\documentclass[12pt]{article}
\usepackage{amsmath,amssymb,epsfig}
\usepackage{epsfig}
\renewcommand{\theequation}{\thesection.\arabic{equation}}
\newlength{\extraspace}
\newlength{\extraspaces}
\newcommand{\be}{\begin{equation}
\addtolength{\abovedisplayskip}{\extraspaces}
\addtolength{\belowdisplayskip}{\extraspaces}
\addtolength{\abovedisplayshortskip}{\extraspace}
\addtolength{\belowdisplayshortskip}{\extraspace}}
\newcommand{\ee}{\end{equation}}
 
\newcommand{\ba}{\begin{eqnarray}
\addtolength{\abovedisplayskip}{\extraspaces}
\addtolength{\belowdisplayskip}{\extraspaces}
\addtolength{\abovedisplayshortskip}{\extraspace}
\addtolength{\belowdisplayshortskip}{\extraspace}}
\newcommand{\ea}{\end{eqnarray}}

\newcommand{\bas}{\begin{eqnarray*}
\addtolength{\abovedisplayskip}{\extraspaces}
\addtolength{\belowdisplayskip}{\extraspaces}
\addtolength{\abovedisplayshortskip}{\extraspace}
\addtolength{\belowdisplayshortskip}{\extraspace}}
\newcommand{\eas}{\end{eqnarray*}}
 
\newcounter{subequation}[equation]
\makeatletter

\expandafter\let\expandafter
\reset@font\csname reset@font\endcsname

\def\subeqnarray{\arraycolsep1pt
    \def\@eqnnum\stepcounter##1{\stepcounter{subequation}%
        {\reset@font\rm(\theequation\alph{subequation})}}
\jot5mm     \eqnarray}

\def\subarray{\arraycolsep1pt
    \def\@eqnnum\stepcounter##1{\stepcounter{subequation}%
        {\reset@font\rm(\alph{subequation})}}
\jot5mm     \eqnarray}

\makeatother

\newcommand{\newappendix}[1]{
\vspace{15mm}
\pagebreak[3]
\addtocounter{section}{1}
\setcounter{equation}{0}
\setcounter{subsection}{0}
\renewcommand{\theequation}{\Alph{section}.\arabic{equation}}
\begin{flushleft}
{\large\bf Appendix \Alph{section}: #1}
\end{flushleft}
\nopagebreak
\medskip
\nopagebreak}

\newcommand{\newsection}[1]{
\vspace{15mm}
\pagebreak[3]
\addtocounter{section}{1}
\setcounter{equation}{0}
\setcounter{subsection}{0}
 
\begin{flushleft}
{\large\bf \thesection. #1}
\end{flushleft}
\nopagebreak
\medskip
\nopagebreak}
 
\newcommand{\newsubsection}[1]{
\vspace{1cm}
\pagebreak[3]
 
\addtocounter{subsection}{1}
\noindent{ \bf \thesection.\arabic{subsection} #1}
\nopagebreak
\vspace{2mm}
\nopagebreak}

\newcommand{\Z}{\mathbb{Z}}
\newcommand{\R}{\mathbb{R}}

\renewcommand{\S}{\mathbb{S}}

\renewcommand{\H}{\mathbb{H}}

\newcommand{\1}{\mbox{1\hspace{-.8ex}1}}
\newcommand{\bra}{\langle}
\newcommand{\ket}{\rangle}
\newcommand{\ra}{\rightarrow}

\newcommand{\up}{\uparrow}
\newcommand{\is}{ &\! =\! & }
\newcommand{\nonum}{\nonumber \\[1.5mm]}
\newcommand{\sspace}{\makebox[1cm]{ }}
\newcommand{\bspace}{\makebox[2cm]{ }}
\newcommand{\nspace}{\!\!\!\!\!\!\!\!\!\!}

\newcommand{\Tr}{{\rm Tr}}

\renewcommand{\th}{{\theta}}
\newcommand{\eps}{\epsilon}
\newcommand{\lb}{\lambda}
\newcommand{\om}{\omega}
\newcommand{\sh}{{\rm sh}}
\newcommand{\ch}{{\rm ch}}

\newcommand{\cD}{{\cal D}}

\newcommand{\cN}{{\cal N}}

\newcommand{\cO}{{\cal O}}

\newcommand{\cS}{{\cal S}}

\newcommand{\mf}{\mathfrak{m}}
\newcommand{\kf}{\mathfrak{k}}
\newcommand{\gf}{\mathfrak{g}}

\newcommand{\n}{{\sc n}}
\renewcommand{\dh}{{\widehat{D}}}

\begin{document}

\begin{titlepage}

\renewcommand{\thefootnote}{\fnsymbol{footnote}}
\begin{flushright}
MPP-2007-34
\end{flushright}
\mbox{}
\vspace{2mm}

\begin{center}
\mbox{{\Large \bf Perturbative and nonperturbative correspondences}}\\[4mm]
\mbox{{\Large \bf between compact and non-compact sigma-models}}\\[4mm]
\vspace{1.3cm}

{{\sc M.~Niedermaier\footnote{Membre du CNRS}, E. Seiler, P. Weisz}}%
\\[4mm]
{\small\sl Laboratoire de Math\'{e}matiques et Physique Th\'{e}orique}\\
{\small\sl CNRS/UMR 6083, Universit\'{e} de Tours}\\
{\small\sl Parc de Grandmont, 37200 Tours, France}
\\[3mm]
{\small\sl Max-Planck-Institut f\"{u}r Physik}\\
{\small\sl F\"ohringer Ring 6}\\
{\small\sl 80805 M\"unchen, Germany}

\vspace{1.4cm}

\end{center}

\begin{quote}
Compact (ferro- and antiferromagnetic) sigma-models 
and noncompact (hyperbolic) sigma-models are compared in a lattice 
formulation in dimensions $d \geq 2$. While the ferro- and 
antiferromagnetic models are essentially equivalent, the
qualitative difference to the noncompact models is highlighted. 
The perturbative and the large $N$ expansions are studied in 
both types of models and are argued to be asymptotic 
expansions on a finite lattice. An exact correspondence between the 
expansion coefficients of the compact and the noncompact models 
is established, for both expansions, valid to all orders on a 
finite lattice. The perturbative one involves flipping the 
sign of the coupling and remains valid in the termwise infinite volume 
limit. The large $N$ correspondence concerns the functional 
dependence on the free propagator and holds directly only in 
finite volume. 
\end{quote} 
\vfill

\setcounter{footnote}{0}
\end{titlepage}


\newsection{Introduction} 

Nonlinear sigma-models with maximally symmetric riemannian 
target spaces naturally come in dual pairs, one compact and the 
other noncompact. The generalized classical spin systems 
associated with both the compact and the noncompact
target spaces  have a variety of applications, see e.g.~\cite{DNS,fg,SZ} 
for the less familiar noncompact models. Since the 
compact models are much better understood and the target manifolds 
can be related to those of the noncompact models by analytic 
continuation, it is natural to try to relate also the quantum 
(or statistical mechanics) properties of the compact to that of 
the noncompact models. For definiteness we consider here as 
target-spaces the $N$-dimensional sphere $\S^N$ and its dual the 
hyperboloid $\H^N$, however the qualitative aspects should be 
the same for other dual pairs. 

The goal of this note is to present two examples of such relations  
for invariant correlation functions.  The first one concerns perturbation
theory, the other the large $N$ expansion. Provided one 
has chosen a formulation where the expansions are 
valid asymptotic expansions, it turns out that knowledge 
of the expansion coefficients in the compact model allows one 
to infer those in the noncompact model. However it is 
crucial that the asymptotic expansions are known to exist 
beforehand in both systems independently. The lattice formulation 
is especially suited to address this and we shall see 
that on a finite lattice the relevant asymptotic expansions 
do exist; for the perturbative one in section 2 below 
and for the large $N$ expansion in a separate paper
\cite{horo}. The perturbative correspondence simply 
involves a sign flip of the coefficients and in a formal 
expansion (dimensional regularization and minimal subtraction) 
has been noted to low orders in \cite{Hought, Hikami} and in the
literature on Riemannian sigma models. Here we present a proof 
of the correspondence to all loop orders, initially on a 
finite lattice; but the termwise infinite volume limit should 
exist on account of the expected lattice counterpart of
David's theorem \cite{Elitzur, David}.  The 
perturbatively defined correlation functions, viewed as functions 
of the lattice points and of $1/\beta$, are thus likewise related 
simply by flipping the sign of $\beta$.  

The large $N$ correspondence is more subtle. In brief the $s$-th order large $N$ 
coefficient $W_{\pm,r}^{(s)}$ of the invariant $2r$-point functions on a finite 
lattice can in the compact ($+$) and in the noncompact ($-$) model be expressed 
in terms of a single functional $X_r^{(s)}[D](\lb)$ of the leading order 
invariant two-point function $D$ -- the same functional for both systems. This 
``large $N$ correspondence'' is computationally useful because the computations 
in the compact model, which do not require gauge fixing, are much simpler. 
The `detour' over finite volume cannot be avoided as the correspondence 
is difficult to interpret directly in the infinite volume limit, see 
\cite{TDlimit}. In addition is important to appreciate that although the gap 
equations are related by flipping the sign of the large $N$ coupling $\lb = 
(N\!+\!1)/\beta$, the leading order propagator $D_-$ in the noncompact model 
behaves very differently as a function of the lattice distance than $D_+$:
while the latter shows exponential decay in the thermodynamic limit, 
the former decreases only with a power law (for $d>2$) or 
increases logarithmically with the distance (for $d=2$).

In a finite volume the gap equations for the compact as well as the noncompact 
model have the same $O(V)$ number of solutions for the (mass)$^2$ parameter; but 
whereas in the compact model there is exactly one positive solution, which is the 
`physical' one, in addition to multiple negative ones, the sign-flipped gap 
equation has only negative solutions of which only one is the right one for 
defining $D_-$. The fundamental criterion by which the relevant solution is 
selected should be the stability of the corresponding saddle point, but we find 
also an alternative characterization selecting the physically relevant saddle 
point of the noncompact model. In the thermodynamic limit, in both cases all the 
negative solutions of the gap equation disappear, therefore the large $N$ 
expanded correlation functions, viewed as functions of the lattice points, are 
not related in any simple way, in particular not merely by flipping the sign of 
$\lb$.

The derivations of the above results also highlight why one cannot 
expect useful correspondences to exist beyond asymptotic expansions. For 
example flipping the sign of the coupling in the (exact) generating 
functionals for invariant correlation functions maps the ferromagnet onto 
the compact antiferromagnet. In contrast, in the perturbative asymptotic 
expansion the sign flip rather relates the ferromagnet to the noncompact 
model, while in the large $N$ expansion the functional dependence on the 
independently defined free propagators gets related. In the latter case 
the relation to the compact model cannot be formulated directly in the 
thermodynamic limit but one can take term by term the thermodynamic limit on 
both sides of the correspondence. This somewhat tricky limit is studied in an 
accompanying paper \cite{TDlimit} in the two-dimensional systems for a number 
of physically interesting invariant quantities. Ultimately we expect the 
fundamental differences between the compact and the noncompact models to be 
rooted in two facts: the presence of a infinite volume mass gap and the absence 
of long range order (for $d \leq 2$) in the compact models and the opposite 
characteristics in the noncompact models for all $d \geq 1$.

The rest of the paper is organized as follows: In section 2 we 
introduce generating functionals for the invariant correlators 
in the three systems considered and establish the perturbative 
correspondence. In section 3 dual formulations of the generating 
functions are introduced, suited for the large $N$ expansions. 
The large $N$ correspondence is shown first from the 
generating functionals, then from the Schwinger-Dyson equations, 
and finally verified at low orders.


\newsection{Invariant correlators and PT correspondence} 

Invariant correlators in the noncompact model have to be defined 
in terms of a gauge fixed generating functional. A gauge fixing where 
one spin is kept fixed turns out to be advantageous. To discuss the 
relation to the compact model we adopt the same gauge fixing there. 
This spin model formulation is also convenient to discuss 
the relation between the exact generating functionals 
and for the proof of the perturbative (PT) correspondence.

\newsubsection{Definitions} 

We begin by setting up the notation and the definitions for the invariant 
correlation functions considered and their generating functionals.
We consider the ${\rm O}(N+1)$ spherical and the ${\rm SO}(1,N)$ 
hyperbolic sigma-models with standard lattice action, defined on 
a hypercubic lattice $\Lambda \subset \Z^d$ of volume 
$V = |\Lambda| = L^d$. The dynamical variables (``spins'') 
will be denoted by $n_x^a$, $x \in \Lambda$, $a =0, \ldots, N$, in both 
cases, and periodic boundary conditions are assumed throughout 
$n_{x + L\hat{\mu}} = n_x$. The constraint is $n\cdot n =1$ in 
both cases, but with different `dot' products; namely $a \cdot b := 
a^0 b^0 + a^1 b^1 + \ldots + a^N b^N =: a^c \delta_{cd} b^d$ in the compact 
model, and $a \cdot b := a^0 b^0 - a^1 b^1 - \ldots - a^N b^N =: 
a^c \eta_{cd} b^d$ in the noncompact model. We shall also use the notation 
$\vec{a} =(a^1, \ldots, a^N)$ for vectors in $\R^{N}$, so that the 
bilinear forms read $a \cdot b = a^0 b^0 \pm \vec{a} \cdot \vec{b}$,
in the two cases. 
Clearly $\S^N = \{ n \in \R^{N+1}\,| \,n\cdot n =1\}$ is the 
$N$-sphere and $\H^N = \{ n \in \R^{1,N}\,|\, n \cdot n =1,\, n^0 >0\}$ 
is the upper half of the two-sheeted $N$-dimensional hyperboloid. 
The invariance groups are ${\rm O}(N+1)$ and ${\rm SO}_0(1,N)$, 
respectively. 

Let us briefly note how $\S^N$ and $\H^N$ are related by symmetric space 
duality (see for instance \cite{kn,helg}). Recall that a symmetric space 
$G/K$ has an involution $\tau$ associated with it such that the Lie 
algebra $\gf$ of $G$ decomposes according to $\gf = \kf \oplus \mf$ as a 
direct sum of vector spaces, where $\kf$ and $\mf$ are even and odd under 
$\tau$, respectively. Furthermore $[\kf,\kf]\subset\kf$, 
$[\kf,\mf]\subset\mf$ and $[\mf,\mf]\subset\kf$. The dual Lie algebra 
$\gf^*$ is then defined as $\gf^*:= \kf \oplus i\mf$ and the corresponding 
dual group $G^*$ is the (simply connected) group whose Lie algebra is 
$\gf^*$. $G^*$ contains a connected subgroup $\widetilde K$ having $\kf$ 
as its Lie algebra; the dual symmetric space $(G/K)^*$ can then be defined 
as $G^*/\widetilde K$. On $\gf$ we have an invariant bilinear form $B$ (in 
our case simply minus the Killing form) which induces a dual bilinear form 
$B^*$ on $\gf^*$. The bilinear forms $B$ and $-B^*$, restricted to $\mf$ 
and $i\mf$, respectively, are positive definite in our case and define the 
metric of the tangent spaces at the origins of the symmetric spaces $G/K$ 
and $(G/K)^*$, respectively. By requiring invariance under $G$ or $G^*$, 
respectively, Riemannian metrics on both symmetric spaces are induced. For 
$G = {\rm SO}(N+1)$, $K = {\rm SO}(N)$, this gives $G^* = \widetilde{{\rm 
SO}}_0(1,N)$ (the universal covering group of ${\rm SO}_0(1,N)$) and 
$(G/K)^*=\widetilde{{\rm SO}}_0(1,N)/\widetilde{{\rm SO}}_0(N)={\rm 
SO}_0(1,N)/{\rm SO}_0(N)$ or $G/K=\S^N$ and $(G/K)^*=\H^N$.

The lattice actions for the two systems are 
\be 
S_{\pm} = \mp \beta \sum_{x,\mu} (n_x \cdot n_{x+\hat{\mu}} -1) = 
\mp \frac{\beta}{2} \sum_x n_x \cdot (\Delta n)_x\,, 
\label{def1}
\end{equation}
where the upper sign refers to the compact model and the lower 
sign to the noncompact model. In the compact model, with the conventions 
adopted in (\ref{wc1}), $\beta >0$ models ferromagnetic behavior while 
$\beta < 0$ models antiferromagnetic behavior. In the noncompact 
model only $\beta >0$ is allowed and the action is unbounded from 
above, $0 \leq S_-[n] < \infty$. The Laplacian is 
$\Delta_{xy} = - \sum_{\mu} [2 \delta_{x,y} - \delta_{x,y + \hat{\mu}} 
- \delta_{x, y - \hat{\mu}}]$, as usual. We write 
\ba 
d\Omega_+(n) \is dn^{N+1}\delta(n\cdot n -1)\,,
\nonum
d\Omega_-(n) \is 2 dn^{N+1} \delta(n\cdot n -1) \theta(n^0)\,,
\label{def2}
\ea
for the invariant measure on $\S^N$ and $\H^N$, respectively.
Further $\delta_{\pm}(n,n')$ is the 
invariant point measure on $\S^N$, $\H^N$, and 
$n^{\uparrow} = (1,0,\ldots,0)$. Note that the 
measure $d\Omega_+(n)$ is normalized while $\H^N$ has 
infinite volume.

In the compact model we consider two generating functionals,
the usual one and a variant with one spin frozen:
\begin{subeqnarray}
\label{wc1}
\nspace \exp W[H] &=& \cN \!\int\! \prod_{x} d\Omega_+(n_x) \exp\Big\{
\!- S_+ + \frac{1}{2} \sum_{x,y} H_{xy}( n_x \cdot n_y -1) \Big\}\,,
\\
\nspace \exp W_+[H] &=& \cN_+ \!\int\! \prod_{x} d\Omega_+(n_x) 
\delta_+(n_{x_0},n^{\uparrow})\,
\exp\Big\{
\!- S_+ + \frac{1}{2} \sum_{x,y} H_{xy}( n_x \cdot n_y -1) \Big\}\,,
\end{subeqnarray}
where $H_{xy} \geq 0$ is a source field and the normalizations $\cN$, 
$\cN_+$ are such 
that $W[0] = 0$. More generally $W_{\!n}[H]$ will denote the generating 
functional with the spin at site $x_0$ fixed by a $\delta(n_{x_0}, n)$ 
insertion. 
In this notation one has $W_+[H] = W_{\!n^{\uparrow}}[H]$ and 
$\int\! d\Omega_+(n) \exp W_{\!n}[H] = \exp W[H]$. In fact 
\be 
W_n[H] = W[H]\,,
\label{wc2}
\end{equation}
for all $n \in \S^N$. This can seen by performing a global 
rotation $n_x \mapsto g n_x =: \tilde{n}_x, \,x \in \Lambda$, 
with $g \in {\rm O}(N\!+\!1)$ chosen such that $g n_{x_0} = 
n^{\uparrow}$, say. The Boltzmann factor will then depend on 
$\tilde{n}_x$, $x \neq x_0$, only and the $d\Omega(\tilde{n}_{x_0}) 
= d\Omega(n_{x_0})$ integration can be performed to give $1$.

In the noncompact model only the fixed spin variant of the 
generating functional is well defined and we write  
\ba 
\label{wnc}
\nspace \exp W_-[H]\! \is \!\cN_- \!\int\! \prod_{x} d\Omega_-(n_x) 
\delta_-(n_{x_0},n^{\uparrow})\,
\exp\Big\{\!
- S_- + \frac{1}{2} \sum_{x,y} H_{xy}( n_x \cdot n_y -1) \Big\}\,,
\ea 
where now $H_{xy} <0$ sources give damping exponentials. 

Partially connected $2r$ point functions are defined by 
\ba 
&& W_{\pm}[H] = \sum_{r \geq 1} \frac{1}{r! \, 2^r}\, 
W_{\pm,r}(x_1,y_1;\ldots;x_r, y_r) \, H_{x_1 y_1} \ldots 
H_{x_r y_r} \,,
\nonum
&& W_{\pm,r}(x_1,y_1;\ldots;x_r, y_r) := h_{x_1 y_1} \ldots h_{x_r y_r}
W_{\pm}[H]\Big|_{H=0}\,,\quad h_{xy} := \frac{\delta}{\delta H_{xy}}\,.
\label{def3}
\ea
In particular $W_{\pm,1}(x,y) = \bra n_x \cdot n_y\ket -1$, 
$W_{\pm,2}(x_1,y_1;x_2, y_2) := \bra n_{x_1} \cdot n_{y_1}  n_{x_2} \cdot
n_{y_2}\ket - \bra n_{x_1} \cdot n_{y_1} \ket \bra n_{x_2} \cdot
n_{y_2}\ket$, where $\bra \;\;\ket$ are the functional averages 
with respect to $\cN_\pm^{-1} e^{-S_{\pm}}$. Note that 
$W_{\pm,r}(\ldots; x,x; \ldots) 
=0$. 

In the above we tacitly assumed that $W_{\pm}[H]$ and the correlation 
functions computed from it do not depend on the site $x_0$ of the frozen 
spin and are translation invariant. We show now that this indeed the case. 
If we momentarily indicate the dependence on the site as $W_{x_0}$ (and 
drop the $\pm$ subscripts) one has trivially 
\be W_{x_0}[\tau_a H] = W_{x_0 
+ a}[H]\,,\quad (\tau_a H)_{xy} = H_{x+a,y+a}\,. \label{def4} 
\end{equation} 
Thus, if $W_{x_0}$ is independent of $x_0$ it is also translation 
invariant. The Boltzmann factors (\ref{wc1}b) and (\ref{wnc}) 
can be viewed as a function on the group via $F(g_0, \ldots, g_s) = f(g_0 
n^{\uparrow}, \ldots , g_s n^{\uparrow})$, where we picked some ordering 
of the sites $x_i, \, i =0,\,1, \ldots , s :=V \!-\!1$, identified 
$n_{x_i}$ with $g_i n^{\uparrow}$, and wrote momentarily $f$ for the 
Boltzmann factor. Then $W_{x_i}$ is of the form 
\ba && \int \!\prod_j d 
\Omega(n_j)\, \delta(n_i, n^{\uparrow}) f(n_0, \ldots, n_s)  = {\rm const} 
\,U_i\,, \nonum && U_i := \int\! \prod_{j \neq i} dg_j \, F(g_0, \ldots, 
g_{i-1}, e, g_{i+1} , \ldots, g_s) \,. 
\label{transl}
\end{eqnarray} 
Using the invariance of $F$ under $g_i \mapsto h^{-1} g_i$ and the 
unimodularity and invariance of the measure $dg$ one verifies that $U_i = 
U_0$ for all $i$.

A peculiarity of the fixed spin gauge is that an invariant $2r$-spin 
correlator $\bra n_{x_1} \cdot n_{y_1} n_{x_2} \cdot n_{y_2} \ldots 
n_{x_r} \cdot n_{y_r} \ket$, can be re-interpreted as a non-invariant 
$2r\!-\!1$-spin correlator. This is manifest when $x_i = x_0$ for some 
$i$; since $n_{x_0}=n^\up$ the correlator then involves only $2r\!-\!1$ 
fluctuating spins. By the above argument any one site can play the role of 
$x_0$, so that picking such a re-interpretation amounts to picking a site 
where $n_{x_i} = n^{\uparrow}$. This feature holds both in the compact and 
in the noncompact models, but initially only on a finite lattice. When the 
site $x_0$ is kept fixed in the interior of the lattice, these specific 
noninvariant $2r-1$ point functions will have a pointwise thermodynamic 
limit in both cases. However the fixed spin averages then do not approach 
limits which can be interpreted in terms of averages without gauge fixing 
or with a translation invariant gauge fixing. This changes if $x_0$ is 
identified with a point on the boundary and moves out to infinity as 
$\Lambda \ra \Z^d$. In this case, in dimensions $d\le 2$ an important 
difference between the compact and noncompact models emerges: while in the 
compact model the Mermin-Wagner theorem assures that the non-invariant 
correlator is equal to its invariant average over ${\rm O}(N+1)$, in the 
noncompact model the non-amenability of ${\rm SO}(1,N)$ prevents this, and 
we find either spontaneous symmetry breaking or divergence of the 
correlators, as discussed in \cite{DNS}.

The goal in the following will be to relate the perturbative and the large $N$ 
expansions in the compact and the noncompact models. 
For the perturbative expansions of the correlation functions, we write 
\be 
W_{\pm,r} \sim \beta^{-r} \sum_{n\geq 0} \beta^{-n}\, w_{\pm,r}^{(n)} \,.
\label{pt0}
\end{equation}
In a large $N$ expansion $\lambda := (N\!+\!1)/\beta$ 
is kept fixed and the coefficient functions $W_{\pm,r}^{(s)}$ in 
\be 
W_{\pm,r} \sim \frac{\lambda^r}{(N+1)^{r-1}}
\sum_{s \geq 0} \frac{1}{(N \! + \! 1)^s} 
W_{\pm,r}^{(s)} \,,
\label{def5}
\end{equation}
are sought.

\newsubsection{Compact antiferromagnet and noncompact model}

In the literature one finds statements, based on formal manipulations of 
functional integrals, suggesting that compact and noncompact models are 
essentially related just by flipping the sign of $\beta$ (in \cite{fg}, 
however, the qualifying remark is added that such a relation is `less 
clear' at the non-perturbative level). Since this simple `flip rule' is 
not completely correct and the relations between the compact model, its 
analytic continuation to $\beta<0$ and the noncompact model are somewhat 
subtle, let us try to clarify the situation. To avoid confusion, in the 
following $\beta$ will always be assumed to be non-negative.

The compact model defined in Eq.~(\ref{def1}) describes for $\beta>0$ a 
ferromagnet, for $\beta<0$ an antiferromagnet, so flipping the sign of 
$\beta$ turns a ferromagnet into an antiferromagnet, not the noncompact 
model. It turns out that one can nevertheless relate the coefficients of 
the perturbation expansion and the large $N$ expansion of the compact 
ferromagnetic and the noncompact model by a sign flip prescription, which 
does, however, not imply any simple relation between the antiferromagnet 
and the noncompact model.

On the simple hypercubic lattices we are considering here, there is 
a transformation between the cases $\beta>0$ and $\beta<0$, based 
on the fact that we can decompose $\Lambda$ into an even and an odd 
sub-lattice
\be
\Lambda=\Lambda_+\cup \Lambda_-\ ,
\end{equation}
where $x\in \Lambda_+$ if and only if $\sum_{i=1}^d x_i$ is even, 
otherwise $x\in \Lambda_-$. Defining 
\be
\epsilon_x:=(-1)^{x_1+\ldots +x_d}\,,
\end{equation}
we can define a map of the configurations by
\be \label{flipbeta}
n_x\mapsto (n_x')=\epsilon_x n_x\,,
\end{equation}
mapping each configuration of the ferromagnet into one of the antiferromagnet 
with the same (correctly normalized) Boltzmann weight. This yields the 
following identity relating the generating functionals of the 
antiferromagnetic and the ferromagnetic systems:
\be
W_{+,-\beta}[H] = W_{+,\beta}[H^{\eps}] +  
\frac{1}{2} \sum_{x y} (H_{xy}^{\eps} -H_{xy}) + {\rm const}\,, 
\label{flipW}
\end{equation} 
where we momentarily indicated the dependence on $\beta$ and set  
$H^\epsilon_{xy}:= \epsilon_x H_{xy}\epsilon_y$. For the invariant two-point 
function this gives 
\be 
\bra n_x\cdot n_y\ket\Big|_{-\beta}= \epsilon_x\epsilon_y \bra n_x\cdot 
n_y\ket\Big|_{\beta}\,,
\end{equation}
and similarly for the higher correlation functions. This implies that the 
physics of the antiferromagnet is essentially the same as for the 
ferromagnet, in particular they have the same mass gap, phase structure 
etc..

The relation between the ferromagnetic ${\rm O}(N\!+\!1)$ and the noncompact 
${\rm SO}(1,N)$ systems (the latter only exists in the ferromagnetic 
version) is, however, more subtle: starting from the ${\rm O}(N\!+\!1)$ 
model we re-parameterize the variables $n_x \in \S^N$ by 
introducing $\beta=:g^{-1}$ and $\vec n_x=:\sqrt{g}\,\vec \pi_x$, 
so that
\be
n_x\cdot n_y=n_x^0n_y^0+g \vec\pi_x\cdot\vec\pi_y=
\sigma_x\sigma_y\sqrt{1-g\vec\pi_x^2}\  
\sqrt{1-g\vec\pi_y^2}+g\vec\pi_x\cdot\vec\pi_y\ , 
\end{equation}
where the $\sigma_x$ are variables taking the values $\pm1$. To obtain 
$W_+[H]$ in this parameterization one has to $(i)$ integrate over all 
$\vec\pi_x$ subject to the constraint $\vec\pi_x^2<\beta$ and 
$(ii)$ sum over 
all $\sigma_x=\pm 1$ except $\sigma_{x_0}$, which is 1. Explicitly
\ba
&&\exp W_+[H]= \widetilde\cN_+ \sum_{\{\sigma_x\}}\delta_{1,\sigma_{x_0}}
\!\int\! \prod_{x} 
d\vec\pi_x\delta(\vec\pi_{x_0}) \theta(|\beta|-\vec\pi_x^2)\ 
(1-g\vec\pi_x^2)^{-1/2}\,\cr
&&\times\exp\left\{-\frac{1}{2g}
\sum_{x,\mu}\left[ \left(\sigma_x\sqrt{1-g\vec\pi_x^2}
-\sigma_{x+\hat\mu}\sqrt{1-g\vec\pi_{x+\hat{\mu}}^2}\right)^2
+g(\vec\pi_x-\vec\pi_{x+\hat\mu})^2\right]\right\}\cr
&&\times
\exp\left\{\frac{1}{2} \sum_{x,y}  
H_{xy} \Big(\sigma_x\sigma_y 
\sqrt{1-g\vec\pi_x^2}\:\sqrt{1-g\vec\pi_y^2}  
+g\vec\pi_x\cdot\vec\pi_y-1 \Big) \right\}\,. 
\label{wc3}
\end{eqnarray}
For $\beta\to+\infty$ (ferromagnet) the term in which all $\sigma_x$ are 
equal dominates (for $\beta\to-\infty$ (antiferromagnet) instead the terms 
with alternating $\sigma_x$ dominate).

On the other hand, we may re-parameterize the noncompact model in a similar 
way by introducing  $\vec \n_x=\sqrt{g}\ \vec \pi_x$, but this time 
\be
n_x\cdot n_y=n_x^0n_y^0-g \vec\pi_x\cdot\vec\pi_y=
\sqrt{1+g\vec\pi_x^2}\  \sqrt{1+g\vec\pi_y^2}-
g\vec\pi_x\cdot\vec\pi_y\ .
\end{equation}

Inserting this into the expression for $W_-[H]$ given in (\ref{wc1}b),
we now obtain
\ba
&&\exp W_-[H]= \widetilde\cN_-\!\int\! \prod_{x}
d\vec\pi_x\delta(\vec\pi_{x_0}) \,
(1+g\vec\pi_{x_0})^{-1/2}\,\cr
&&\times\exp\left\{\frac{1}{2g}\sum_{x,\mu}\left[
\left(\sqrt{1+g\vec\pi_x^2}-\sqrt{1+g\vec\pi_{x+\hat\mu}}\right)^2
-g(\vec\pi_x-\vec\pi_{x+\hat\mu})^2\right]\right\}\cr
&&\times
\exp\left\{\frac{1}{2} \sum_{x,y}
H_{xy} \Big(\sqrt{1+g\vec\pi_x^2}\:\sqrt{1+g\vec\pi_y^2}
-g\vec\pi_x\cdot\vec\pi_y-1 \Big) \right\}\,. 
\label{wnc2}
\end{eqnarray}
Comparing now Eqs (\ref{wc3}) and (\ref{wnc2}), we see that the partition 
function of the noncompact model is obtained from the one of the compact 
one by $(i)$ flipping the sign of $\beta$, $(ii)$ dropping all terms except 
the one with all $\sigma_x=1$ and $(iii)$ omitting the $\theta$ functions 
restricting the domain of integration.

So it is clear that the analytic continuation of the compact system to 
negative $\beta$ (the antiferromagnet) is {\it not} equivalent to the 
noncompact ${\rm SO}(1,N)$ system, as one might infer from 
a cursory reading of the literature (for instance \cite{fg}). This 
non-equivalence will be shown even more manifestly below. Within the 
framework of the Schwinger-Dyson Equations used in section 3.5 a 
noteworthy consequence is that even the exact SD equations do not 
determine their solution uniquely.

In the following sections we will see, however, that nevertheless the 
perturbation expansions in a finite volume of the ferromagnetic compact 
and the noncompact model are related simply by flipping the sign of 
$\beta$ (but without the introduction of the sign factor $\epsilon_x$ as 
in the antiferromagnet); furthermore we will derive a relation between 
the $1/N$ expansions of the two models.


\newsubsection{Perturbative correspondence}

We now consider the perturbative expansions (\ref{pt0}) of the correlation 
functions. On a finite lattice it turns out the perturbative series in 
the compact ferromagnetic and in the noncompact models are asymptotic 
expansions which are simply related by a sign flip in the coupling, 
$\beta \mapsto -\beta$; the perturbation expansion of the noncompact 
model is, however, not equal to that of the antiferromagnet, as one might 
guess from this.
 
Concretely
\be 
w_{-,r}^{(n)} = (-)^{r+n} w_{+,r}^{(n)}\,,\quad \forall \,r \geq 1, \, n \geq
0\,,
\label{pt1}
\end{equation}
for the coefficients (\ref{pt0}). The sign flip rule was stated explicitly 
by Hikami \cite{Hikami} (in a formal continuum expansion based on 
dimensional regularization) but was presumably known to other authors in 
special cases and at low orders, e.g.~\cite{Hought}. It has also been 
observed in the perturbative beta functional of Riemannian sigma 
models (using dimensional regularization and minimal subtraction) and can 
readily be verified to all loop orders in this framework. But to the best 
of our knowledge no general proof is available in print, so we present a 
simple proof here.

We set $Z_{\pm}[H] := \exp W_{\pm}[H]$ and define
\newpage
\ba
&&Z^{\rm PT}_+[H](\beta):=\int\! \prod_{x}
d\vec\pi_x\delta(\vec\pi_{x_0})\theta(\beta-\vec\pi_x^2) \,
(1-g\vec\pi_{x}^2)^{-1/2}\,\cr
&&\times\exp\left\{-\frac{\beta}{2}
\sum_{x,\mu}\left[\left(\sqrt{1-g\vec\pi_x^2}-\sqrt{1-g
\vec\pi_{x+\hat{\mu}}^2}\right)^2+g(\vec\pi_x-\vec\pi_{x+\hat\mu})^2\right]
\right\}
\nonum
&&\times
\exp\left\{\frac{1}{2} \sum_{x,y}
H_{xy}\Big(\sqrt{1-g\vec\pi_x^2}\:\sqrt{1-g\vec\pi_y^2}
+g\vec\pi_x\cdot\vec\pi_y-1 \Big) \right\}\,,
\label{wnc3}
\end{eqnarray}
and
\ba
&&Z^{\rm PT}_-[H](\beta):=\!\int\! \prod_{x}
d\vec\pi_x\delta(\vec\pi_{x_0})\theta(\beta-\vec\pi_x^2) \,
(1+g\vec\pi_{x}^2)^{-1/2}\,\cr
&&\times\exp\left\{\frac{\beta}{2}
\sum_{x, \mu} \left[\left(\sqrt{1+g\vec\pi_x^2}-\sqrt{1+g
\vec\pi_{x+\hat{\mu}}^2}\right)^2- g(\vec\pi_x-\vec\pi_{x+\hat\mu})^2\right]
\right\}
\nonum
&&\times 
\exp\left\{\frac{1}{2} \sum_{x,y}
H_{xy} \Big(\sqrt{1+g\vec\pi_x^2}\:\sqrt{1+g\vec\pi_y^2}
-g\vec\pi_x\cdot\vec\pi_y-1 \Big) \right\}\,.
\label{wc4}
\end{eqnarray} 
$Z_+^{\rm PT}[H]$ is, up to a multiplicative constant, just the term of 
$Z_+[H]$ (Eq.(\ref{wc3})) in which $\sigma_x=1$, for all $x$. 
$Z_-^{\rm PT}[H]$ differs from $Z_-[H]$  in Eq.~(\ref{wnc2}) by the 
presence of the $\theta$ functions and a multiplicative constant. 
We now state the

{\bf Result:}
\be 
Z_{\pm}[H] = c_\pm(\beta)Z_{\pm}^{\rm PT}[H](\beta)+
 O(\beta^{-\infty}) \,,
\label{pt2}
\end{equation}
where $c_\pm(\beta)$ are normalization constants independent of $H$.
Further the $Z_{\pm}^{\rm PT}[H](\beta)$ have asymptotic expansions 
of the form 
\be 
Z_{\pm}^{\rm PT}[H](\beta) \sim \sum_{n=0}^\infty a_n[H] (\pm 
\beta)^{-n}\,,
\label{pt3}
\end{equation}
with {\it identical} coefficients $a_n[H]$ in both cases.  
  
The existence of the asymptotic expansions (\ref{pt3}) is a standard result 
of Laplace's method in asymptotic analysis (see for instance \cite{copson,
murray}). The relation (\ref{pt2}) expresses the irrelevance of 
boundaries away from the maximum as well as of the terms in which the 
$\sigma_x$ are not all equal. It follows from two facts:
 
Fact 1: On a finite lattice in the compact ferromagnetic model all 
terms in $Z_+[H]$ except the one with $\sigma_x=1$ are $O(\beta^{-\infty})$.

Fact 2: On a finite lattice in the noncompact model for any 
$x_0\in\Lambda$ and $\beta\ge \beta_0>0$ the expectation value
\ba
\langle\theta(\vec n_z^2-1)\rangle &: =&
\frac{1}{Z_-[H]} \int\! \prod_x d\Omega_{-}(n_x)\, \delta(n_{x_0},n^\up)\, 
\theta(\vec{n}^2_z-1)
\nonum
&& \times \exp\Big\{\!-\!S_-+\frac{1}{2}\sum_{x,y}H_{xy}(n_x\cdot n_y-1)\Big\}\,,
\end{eqnarray}
is bounded by $a \exp(-b\beta)$ with $a,b>0$ .

Corollary: There are constants $a_\pm,b_\pm>0$ such that for 
$\beta>\beta_0$
\be
\left|\frac
{Z_\pm^{\rm PT}[H](\beta) \beta^{- N \frac{V-1}{2}}}   
{Z_\pm[H](\beta)}-1\right|
\le a_\pm \exp(-b_\pm\beta)\,.
\end{equation}
Both facts have essentially the same origin: the fact that the `energy' in 
the omitted contributions is exponentially small. We give a detailed 
proof only of Fact 2 and add some comments on the proof of Fact 1. 

{\it Proof of Fact 2:} We do the integrations in the following way:
first we integrate over all configurations having a fixed value $s$ of the
action and afterwards over $s$. Thus we can write
\be
Z_-[H](\beta)= \int\! ds \,e^{-\beta s} \rho_H(s)\ ,
\end{equation}
with some non-negative density function $\rho_H$. If we then can show that
for any configuration violating the field cutoff the action is larger than
$s_0$ (`energy bound'), it will follow that
\be
Z_-[H](\beta)\,\langle \theta (\vec n_x^2-1)\rangle
\le \int\! ds \, e^{-\beta s} \rho_H(s) \theta(s-s_0)\, .
\label{bound}
\end{equation}
The right hand side is easily seen to be 
$\le \exp(-\frac{\beta}{2}s_0)\times Z_-[H](\beta/2)$. 
By standard asymptotic analysis (see for instance \cite{copson, murray}) 
it follows that $Z_-[H](\beta)/Z_-[H](\beta/2)$ goes to a finite limit as 
$\beta\to \infty$, and therefore
\be
\frac{Z_-[H](\beta/2)}{Z_-[H](\beta)}\le a\,,
\end{equation}
for $\beta\ge \beta_0$. So Eq.~(\ref{bound}) implies Fact 2.

It remains to prove the `energy bound'. Since we are using the fixed
spin gauge, we have $n_{x_0}=n^\uparrow$. Then by assumption there
is a site $x_1$ with $n^0_{x_1}\ge \sqrt{2}$. Choose a path P of
length $|P|\le V$ from $x_0$ to $x_1$ and denote the part of the action   
corresponding to the path P by $s_P$. We claim that $s_P$ is minimized by
moving in equal steps along the geodesic from $n_{x_0}=n^\uparrow$ to
$n_{x_1}$, which implies
\be
s\ge s_P\ge \frac{c}{|P|}\ge \frac{c}{V}\,,
\end{equation}
with some constant $c$.

To see the lower bound on $s_P$ it is sufficient to consider a path of
three points, since the minimization condition is local. Thus we only have
to minimize $e(n):=n_0\cdot n+n\cdot n_1$ over $n$ and show that the
minimum is assumed for $n$ being the midpoint of the geodesic in $\H^N$ 
from $n_0$ to $n_1$.
Choosing coordinates such that $n_0=\n^\uparrow$ and $n_1=(\ch\theta_1,
\sh\theta_1,0,\ldots0)$  with $\theta_1\ge 0$ we find
\be
e(n)=n^0 (1+\ch\theta_1)-n^1\ \sh\theta_1\ \ge n^0
(1+\ch\theta_1)-\sqrt{(n^0)^2-1}\ \sh\theta_1\ .
\end{equation}
Putting $n^0=\ch\theta$ this becomes
\be
e(n)\ge \ch\theta+\ch(\theta_1-\theta)\ge 2 \ch\frac{\theta_1}{2}\ ,
\end{equation}
using the convexity of $\ch$ function in the last step. It follows that
\be
s_P\ge |P|\left(\ch(\theta_1/|P|)-1\right) \ge \frac
{\theta_1^2}{2|P|}\ge \frac {c}{V}\,,
\end{equation}
with $c= ({\rm arccosh} \sqrt{2})^2/2$. The proof of Fact 2 is now 
complete with $a$ as above and $b= c/V$.

To show Fact 1 one proceeds essentially in the same way; only the energy 
bound is more easily derived in a slightly different way: again it suffices 
to consider three spins $n_0, n, n_1$ with the energy
\be
e(n)=n\cdot n_0+ n\cdot n_1\,,
\end{equation}
direct minimization over $n$ yields 
\be
n= \lb (n_0+n_1)\,, 
\end{equation}
with $\lb$ fixed by the requirement $n^2=1$. So $n$ has to lie on the 
great circle connecting $n_0$ and $n_1$. As before we conclude from this 
that the minimal energy for a path is obtained by moving along the 
shorter part of a great circle (geodesic) connecting $n_0$ to $n_1$ in 
equal steps.

The corollary as well as the main result are obvious consequences.

Thus we have learned that the perturbative expansions of the compact 
ferromagnetic model and the noncompact model are related just by 
flipping the sign of $\beta$, even though doing the same sign flip to the 
full model leads to a completely different model -- the antiferromagnet.

The relation between the asymptotic expansions in $1/\beta$ for the 
compact and noncompact models is given in (\ref{pt1}).  For the two point 
function this means for instance
\be 
w_{-,1}^{(n)}= (-1)^{n+1} w_{+,1}^{(n)}\ .   
\end{equation}
This should be contrasted with the relations between the perturbation 
expansions of the two-point function between the ferromagnetic and 
anti ferromagnetic compact models, which follows from (\ref{flipbeta})
\be
w_{-,1}^{(n)}(x,y)|_{\rm antiferro}= \epsilon_x\epsilon_y 
w_{-,1}^{(n)}(x,y)|_{\rm ferro}\ . 
\end{equation}
It is instructive to see what this means for the lowest order (tree graph)
asymptotics of the two point function:
\ba
\bra n_x\cdot n_y\ket& = & 1+\frac{1}{\beta} w_{+,1}^{(0)}(x,y)+O(\beta^{-2}) 
\quad\quad\quad\quad\,\, {\rm ferromagnet} 
\nonum
\bra n_x\cdot n_y\ket& = & \left(1+\frac{1}{\beta} w_{+,1}^{(0)}(x,y)\right)
\epsilon_x\epsilon_y +O(\beta^{-2})\quad {\rm antiferromagnet} 
\\[2mm]
\bra n_x\cdot n_y\ket& = & 1-\frac{1}{\beta} w_{+,1}^{(0)}(x,y)+O(\beta^{-2}) 
\quad\quad\quad\quad\,\, {\rm noncompact\  model} \ ,
\nonumber
\end{eqnarray}
where 
\be
w_{+,1}^{(0)}(x,y)=\frac{N}{V}\sum_{p\neq 0}\frac{\cos p\!\cdot\! 
(x-y) -1}{E_p}\,,
\end{equation}
with $E_p$ as in Eq. (\ref{prop2}) below.

Note that although $W_+[H]$ for the ferromagnet and the 
antiferromagnet are related by flipping  the sign of $\beta$, 
this is no longer true for the asymptotic expansions, even 
in finite volume. Rather the sign flipped perturbative expansion 
is that of the noncompact model.


\newsection{Dual formulations and large $N$ correspondence}

The goal of this section is to establish a relation between the
large $N$ expansions of invariant correlators in the compact and the
noncompact sigma-models. In the compact model the large $N$ 
expansion is based on the familiar dual formulation, which here 
has to be modified due to the fixed spin. In the noncompact model
the duality transformation is ill defined but by performing 
Gaussian integrations in horospherical coordinates, one can 
obtain a well-defined counterpart of the dual formulation.

\newsubsection {Noncompact generating functional via horospherical 
coordinates} 

Before turning to the expansions proper we present here an 
exact rewriting of $W_-[H]$ based on a partial evaluation 
where the $(\dim \H_N)^V$ integrations are reduced to $V$
integrations. The point of departure is the fact that the hyperboloid 
$\H_N$ admits an alternative parameterization in terms of 
so-called horospherical coordinates. These arise 
naturally from the Iwasawa decomposition of ${\rm SO}_0(1,N)$. 
Here it suffices to note the relation to the hyperbolic spins
\be 
n^0 = \ch \th + \frac{1}{2} \sum_{i=1}^{N-1}t_i^2 e^{-\th}\,,
\quad 
n^1 = \sh \th + \frac{1}{2} \sum_{i=1}^{N-1}t_i^2 e^{-\th}\,,
\quad n^i = e^{-\th} t_{i-1},\;\;i = 2,\ldots, N\,,
\label{def7}
\end{equation} 
 and that $(\th, t_1,\ldots ,t_{N-1}) 
\in \R^N$  defines a globally valid system of coordinates.
It is convenient to write $\vec{t} = (t_1, \ldots, 
t_{N-1})$ 
and $\vec{t}\cdot \vec{t}' = t_1 t_1' + \ldots t_{N-1} t_{N-1}'$. 
For the dot product of two spins $n_x,\,n_y \in \H_N$ this gives 
\be 
n_x \cdot n_y = \ch(\th_x - \th_y) + \frac{1}{2} (\vec{t_x} - \vec{t}_y)^2 
e^{-\th_x - \th_y}\,,
\label{def8}
\end{equation} 
and for the measure in (\ref{def2}) 
\be 
d\Omega_-(n) = e^{-\th(N-1)} d\th \,dt_1\ldots dt_{N-1} = 
e^{-\th(N-1)} d\th \,d \vec{t}\,. 
\label{def9}
\end{equation}
The key advantage of horospherical coordinates is manifest from 
(\ref{def7}), (\ref{def8}): for a quadratic action of the form $S_-$ 
in (\ref{def1}) the integrations over the $\vec{t}$ variables are 
Gaussian and can be performed without approximations. We refer to 
\cite{horo} for the derivation and only note the result 
\ba
\label{Wa1} 
&& \exp W_-[H]= \exp\Big\{ - \frac{1}{2} \sum_{x,y} H_{xy} \Big\} 
\cN \int_{\cD(H)} \prod_{x \neq x_0} da_x 
\nonum 
&& \quad \times 
\exp\Big\{ - \frac{N+1}{2} \Tr \ln \widehat{A} - \frac{\beta}{2} 
\sum_{x \neq x_0} a_x + \frac{\beta}{2}  (\widetilde{A}^{-1})_{x_0x_0}^{-1}
\Big\}\,.
\end{eqnarray} 
Here 
\be 
\label{Wa2}
A_{xy} = - \Delta_{xy} + \frac{1}{\beta} H_{xy} + \delta_{xy} a_x 
= \widetilde{A}_{xy} + a_{x_0} \delta_{x x_0} \delta_{xy}\,, 
\end{equation}
and  $\widehat{A}_{xy}$ is the matrix obtained from $A_{xy}$ by 
omitting the $x_0$-th row and column. The domain $\cD(H)$ is an 
algebraic variety described by 
\be
\cD(H) =  \{ a \in [-2 d, \infty]^{V-1} \,|\, 
\det \widehat{A} >0\,,\;\; \widehat{A}\; 
\mbox{positive semidefinite} \}\,.
\label{Wa3}
\end{equation} 
We also anticipate from \cite{horo} the following 

{\bf Result:} The correlation functions $W_{-,r}$ admit an 
asymptotic expansion of the form (\ref{def5}),
whose (uniquely defined) expansion coefficients coincide with
those defined by the Laplace expansion of (\ref{Wa1}) where 
$\cD(H)$ has been replaced by $\R^{V-1}$. In turn these coefficients 
coincide with those of the formal large $N$ expansion of 
\ba
\label{wdnc}
\exp W^d_-[H] \is  \exp\Big\{-\frac{1}{2} \sum_{x,y} H_{xy} \Big\} 
\,\cN\! \int\! \prod_{x \neq x_0} d\alpha_x 
\exp\Big\{ - (N\!+\!1) S_-[\alpha,H]\Big\}\,,
\nonum
S_-[\alpha,H] \is \frac{1}{2} 
\Tr \ln \widehat{A} + i \sum_{x \neq x_0} \alpha_x 
-\frac{1}{2\lb} (\widetilde{A}^{-1})_{x_0x_0}^{-1} \,,
\end{eqnarray} 
where $a_x$  corresponds to $2 i \lb \alpha_x$. However 
\be 
\label{ineq}
W_-[H] \neq W_-^d[H]\,.
\end{equation}
Heuristically the origin of the `dual' generating functional 
$W_-^d[H]$ can be understood by a dualization of the 
`spatial'  spin components $\vec{n}_x,\,x \in \Lambda$,  
and a formal contour deformation in field space \cite{TDlimit}. The formal 
nature of the latter is responsible for (\ref{ineq}), although
the large $N$ expansion coefficients of $W_-[H]$ are correctly 
reproduced.


\newsubsection{Dual formulation: compact model} 

In the compact model the counterpart of (\ref{wdnc}) is obtained along the 
familiar lines: one  first implements the constraints $n_x \!\cdot \!n_x =1$ 
via a Lagrange multiplier field and then performs the Gaussian integrations. 
Of course in the compact model no gauge-fixing is required and the result is 
well-known. For the purposes of comparing with the noncompact model, however, 
we want to perform the dualization here in the fixed spin gauge, 
i.e.~for $W_+[H]$, in which case certain modifications occur. 
The resulting dual generating functional $W^d_+[H]$ is an 
exact rewriting of the original one: $W[H] = W_+[H] = W_+^d[H]$.

In the fixed spin gauge Gaussian integrals of the following 
form arise
\begin{subeqnarray}  
\label{d1} 
&& \int \prod_x d\phi_x \delta(\phi_{x_0})\, 
\exp\Big\{ - \frac{1}{2} \sum_{x,y} \phi_x A_{xy} \phi_y 
+ \sum_x J_x \phi_x\Big\}   
\nonum
&& \quad = (2 \pi)^{\frac{V-1}{2}} (\det \widehat{A})^{-1/2} 
\exp \Big\{ \frac{1}{2} \sum_{x,y} J_x (\widehat{A}^{-1})_{xy} J_y
\Big\}\,, 
\\
&& \int \prod_{x\neq x_0} d\phi_x 
\exp\Big\{ - \frac{1}{2} \sum_{x,y} \phi_x A_{xy} \phi_y \Big\} 
\nonum
&& \quad = (2 \pi)^{\frac{V-1}{2}} (\det \widehat{A})^{-1/2} 
\exp \Big\{ -\frac{\phi_{x_0}^2}{2(A^{-1})_{x_0x_0}}  
\Big\}\,, 
\end{subeqnarray}
for a real field $\phi_x, \, x \in \Lambda$, and a symmetric 
invertible $L^d\times L^d$ matrix $A$, for which we assume 
that $(A^{-1})_{x_0x_0} \neq 0$.  
Here $\widehat{A}$ is the $(L^d -1)\times (L^d - 1)$  matrix arising from 
$A$ by deleting its $x_0$-th row and column. The inverse of $\widehat{A}$ 
can be expressed in terms of the inverse of $A$ via 
\be 
(\widehat{A}^{-1})_{xy} = (A^{-1})_{xy} - 
\frac{(A^{-1})_{xx_0}(A^{-1})_{yx_0}}{(A^{-1})_{x_0x_0}}\,.
\label{d2}
\end{equation}    
This equation makes sense a priori for $x,y,\neq x_0$, but we may 
trivially extend the matrix $\widehat{A}^{-1}$ to a $L^d\times L^d$ matrix
by setting $([\widehat{A}^{-1}]_{\rm ext})_{x_0 x} =0$ in accordance with 
Eq.~(\ref{d2}). Then $A$ and $[\widehat{A}^{-1}]_{\rm ext}$ are of course 
not inverse to each other
\be 
\sum_z A_{x z} ([\widehat{A}^{-1}]_{\rm ext})_{zy} = \delta_{xy} - 
\delta_{x x_0} \frac{(A^{-1})_{x_0 y}}{(A^{-1})_{x_0x_0}}\,.
\label{d3}
\end{equation}
To verify (\ref{d1}) one writes $\sum_{x,y} \phi_x A_{xy} \phi_y 
= \sum_{x,y \neq x_0} \phi_x A_{xy} \phi_y + 2 \phi_{x_0} 
\sum_{x\neq x_0} A_{x_0 x} \phi_x + \phi_{x_0}^2 A_{x_0x_0}$. To get 
(\ref{d1}a) 
one first does the trivial $\phi_{x_0}$ integration. To get (\ref{d1}b) 
one uses (\ref{d3}). The determinant of $\widehat{A}$ is related to 
that of $A$ by 
\be 
\det A = \frac{\det \widehat{A}}{ (A^{-1})_{x_0x_0}} \,.
\label{d4}
\end{equation}
Often a term in the $x_0$-th matrix element on the diagonal of $A$ has 
to be split off according to $A_{xy} = \widetilde{A}_{xy} - c 
\delta_{xy} \delta_{x_0 x}$. In this case the inverse of $A$ is related 
to the inverse of $\widetilde{A}$ by 
\be 
(A^{-1})_{xy} = (\widetilde{A}^{-1})_{xy} + 
\frac{c}{ 1 - c (\widetilde{A}^{-1})_{x_0x_0}} 
(\widetilde{A}^{-1})_{x x_0}(\widetilde{A}^{-1})_{y x_0}\,.
\label{d6}
\end{equation} 
In particular $A_{x_0x_0} - (A^{-1})_{x_0x_0}^{-1} = \widetilde{A}_{x_0x_0} - 
(\widetilde{A}^{-1})_{x_0x_0}^{-1}$ and 
\be
\frac{1}{(A^{-1})_{x_0 x_0}} = -c + \frac{1}{(\widetilde{A}^{-1})_{x_0x_0}}\,,
\sspace \frac{(A^{-1})_{xx_0}}{(A^{-1})_{x_0x_0}} = 
\frac{(\widetilde{A}^{-1})_{xx_0}}{(\widetilde{A}^{-1})_{x_0x_0}}\,.
\label{d7}
\end{equation} 

With these preparations at hand the dualization of $W_+[H]$ 
is straightforward. It is instructive to start from $W[H]$, single out 
one spin, $n_{x_0}$, and to postpone its dualization. 
Inserting 
\be 
\delta(n_x^2 -1) = (N+1)\int \frac{d\alpha_x}{2\pi} 
\exp\{ - i (N+1) \alpha_x (n_x^2 -1) \}\,\quad \mbox{for}\quad x\neq x_0\,,
\label{cd1}
\end{equation} 
the interchange in the order of integrations can be justified 
and the relevant Gaussian is of the form (\ref{d1}b) 
\ba
&& \int \prod_{x \neq x_0} d n_x \exp\Big\{ - \frac{1}{2}  \frac{N+1}{\lb}
\sum_{x,y} n_x \widetilde{A}_{xy} n_y \Big\}
\nonum
&& \quad = {\rm Const} \, (\det \widehat{A})^{-(N+1)/2} 
\exp\Big\{ - \frac{1}{2}  \frac{N+1}{\lb} 
(\widetilde{A}^{-1})_{x_0x_0}^{-1}n_{x_0}^2 \Big\}\,,
\nonum
&& \widetilde{A}_{xy} = - \Delta_{xy} + 2 i \lb \alpha_x 
(1-\delta_{x_0 x}) \delta_{xy} - \frac{\lb}{N+1} H_{xy}\,. 
\label{cd2}
\end{eqnarray}
This gives for the dual generating functionals 
\ba
\label{cd3}
\exp W^d[H] \is \int \! dn_{x_0} \delta(n_{x_0}^2 -1) 
\exp W^d_{\!n_{\!x_0}}[H]\,,
\nonum
\exp W^d_{\!n_{\!x_0}}[H] \is \exp\Big\{-\frac{1}{2} \sum_{x,y} H_{xy} \Big\}
\cN\!\int\! \prod_{x \neq x_0} d\alpha_x \exp\Big\{\! - 
(N\!+\!1) S_{\!n_{x_0}}[\alpha,H]\Big\}\,,
\nonum
S_{\!n_{\!x_0}}[\alpha,H] \is \frac{1}{2} 
\Tr \ln \widehat{A} - i \sum_{x \neq x_0} \alpha_x 
+\frac{1}{2\lb} n_{x_0}^2 (\widetilde{A}^{-1})_{x_0x_0}^{-1} \ .
\end{eqnarray} 
Here we used the fact $\widehat{\widetilde A}=\widehat A$;  
$W^d_{\!n_{\!x_0}}[H]$ is the generating functional in the 
fixed spin gauge and $S_{\!n_{\!x_0}}[\alpha,H]$ is its dual 
action. We shall also write $S_+[\alpha,H]$ for $S_{n^{\uparrow}}[\alpha,H]$ 
and $W_{+}^d[H]$ for $W_{n^{\uparrow}}^d[H]$. In contrast to the 
noncompact model one has 
\be 
W_+^d[H] = W_+[H]\,,\sspace  W^d[H] = W[H]\,,
\label{equ}
\end{equation}  
and also $W_+[H] = W[H]$, by (\ref{wc2}).  

As a check on (\ref{cd3}) one can verify that by dualizing also 
the last spin one recovers the familiar expressions. Indeed, 
\ba 
\label{cd4}
&& \int\! dn_{x_0} \delta(n_{x_0}^2 -1) \exp\Big\{ - \frac{(N\!+\!1)}{2\lb} 
(\widetilde{A}^{-1})_{x_0x_0}^{-1} n_{x_0}^2\Big\} 
\nonum
&& = \Big(\frac{2\pi\lb}{N\!+\!1}\Big)^{(N-1)/2} 
\int d\alpha_{x_0} e^{i(N+1) \alpha_{x_0}} 
\Big[ (\widetilde{A}^{-1})_{x_0x_0}^{-1} + 2 i \lb \alpha_{x_0}
  \Big]^{-(N+1)/2}\,.
\end{eqnarray}   
Using now (\ref{d6}) for $c = -2 i \lb \alpha_{x_0}$ and 
(\ref{d4}) one obtains 
\ba 
\label{cd5}
\exp W[H] \is \cN \int\! \prod_x d \alpha_x 
\exp\Big\{ - \frac{N+1}{2} \Tr \ln A + i \sum_x \alpha_x \Big\}\,,
\nonum
A_{xy} \is - \Delta_{xy} + 2 i \lb \alpha_x \delta_{xy} 
- \frac{\lb}{N+1} H_{xy}\,,
\end{eqnarray} 
as required.

\newsubsection{Finite volume mass gap and basic propagators} 

The saddle point expansions of $W_{\pm}^d[H]$ define to leading order 
the gap equations and the invariant two-point functions. In both 
quantities the dependence on the site $x_0$ of the fixed spin drops out,
see section 3.6. The result for the leading order two-point function 
can be written in the form 
\be 
\lb W_{\pm,1}^{(0)}(x,y) = \pm \lb D_{\pm}(x-y) -1 = 
\bra n_x \cdot n_y \ket\big|_{N = \infty} -1 \,,
\label{prop1}
\end{equation}
where $D_{\pm}(x-y)$ is the basic propagator. For the compact model it 
has the well-known structure $D_+(x) = D(x)|_{\om = \om_+}$,
where 
\be 
D(x) = \frac{1}{V} \sum_p \frac{e^{ip \cdot x}}{E_p + \om}\,,
\label{prop2}
\end{equation}
with the sum over all $p = \frac{2\pi}{L}(n_1,\ldots, n_d)$,
$n_i = 0,1,\ldots, L\!-\!1$, and $E_p := 2 d - 2 \sum_\mu \cos(p \cdot
\hat{\mu})$. Further $\om_+ = \om_+(\lb,V)$ is the dynamically 
generated finite volume mass$^2$ term determined by the gap equation 
$D(0) = 1/\lb$. A subtlety is due to the fact that the equation for 
$\om_+(\lb,V)$ is an algebraic equation of degree $O(V)$ and therefore has 
$O(V)$ solutions. Among those solutions there is always a unique positive one, 
but it is not obvious that this is always the physically relevant one: in 
dimension $d\ge 3$ for weak coupling there is spontaneous symmetry breaking 
and it has been found in \cite{mukaida, dsw} that at least in the 
determination of the constrained effective potential one has to take a 
negative solution of the gap equation.  Here we want to eliminate these 
complications by always working in the `strong coupling' regime, which we will 
now specify. 

In the infinite volume limit the gap equation becomes
\be
1 = \lb\! \int\! \frac{d^dp}{(2\pi)^d} \frac{1}{E_p+\omega}\,, 
\end{equation}
and it has no negative solutions. It has a unique positive solution 
for $\lb\ge\lb_*(d)$, where $\lb_*(2)=0$ and $\lb_*(d)>0$ for $d\ge 3$ 
($\lb_*(d)$ is known as the critical coupling of the spherical model).
From now on we will assume $\lb > \lb_*(d)$ and always require 
$\om_+(\lb,V)>0$. In this case it is possible to show that the finite volume 
corrections $\om_+(\lb, V)- \om_+(\lb, \infty)$  are exponentially small.

In the noncompact model one has similarly $D_-(x) = D(x)|_{\om = \om_-}$, 
where $\om_-$ is a solution of the gap equation $\lb D(0) = -1$,
see Eqs.~(\ref{gap1}), (\ref{gap2}) in section 3.6. Clearly analytically 
continuing the multivalued algebraic function $\om_+(-\lb,V)$ to negative
values of $\lb$ and setting $\om_-(\lb, V): = \om_+(-\lb,V)$ will produce
solutions of this gap equation. But again there are $O(V)$ such solutions, all 
of 
which are negative. To see this it is convenient to split off the zero momentum 
mode and to write 
\be 
D(0) = \frac{1}{V \om} + f(\om)\,,\quad 
f(\om) = \frac{1}{L^d}\sum_{0 \neq n \in [0,L-1]^d} \frac{1}{\om + \om_n}\,,
\label{prop3}
\end{equation} 
where $\om_n = E_{2\pi n/L}\,,\;n =(n_1,\ldots,n_d)$. Here 
$f(\om)$ is a meromorphic functions with simple poles along the 
negative real axis. The first pole is at $\om = -4\sin^2(\pi/L)$, 
the others are at approximately integer multiples thereof. Specifically 
\be 
\om_n = - 4(\pi/L)^2 (n_1^2 + \ldots + n_d^2) + O(1/L^4)\,.
\label{prop4}
\end{equation}
In terms of the function $f$ the gap equations $\pm \lb D(0) =1$ read 
\be 
\lb f(w) = \pm 1 - \frac{\lb}{L^d \om} \,.
\label{prop5}
\end{equation} 
In the compact model (upper sign) is has a unique solution 
$\om_+(\lb,L)>0$ for any given $\lb > \lb_*(d)$; all the other $O(V)$ 
solutions are negative. In the noncompact case there are only $O(V)$ 
negative solutions, one in each interval of the $\om$ axis between 
consecutive poles of $f$. 

To obtain an asymptotic expansion in $1/N$ one has to pick the solution 
corresponding to the absolute maximum of the dual Boltzmann factor. But there 
is a simpler way to do this selection: if the large $N$ expansion is to yield 
an asymptotic expansion of the invariant two-point function, we have to demand 
that $-\lb D(x) \geq 1$ for all $x$. Since $\bra n_x \cdot n_y\ket \geq 1$ in 
the noncompact model, this is a necessary condition for the interpretation of 
$D_-(x-y) = D(x-y)|_{\om = \om_-}$ as the leading order two-point function. 
This condition fixes the solution to lie in the interval $\om \in (-4 \sin^2 
\pi/L, 0)$. Indeed, $-\lb D(x) \geq 1$ implies
\ba
\sum_x (1 - \cos q\! \cdot\! x) (-\lb D(x)) &\geq& V\,,\quad 
q \neq 0\,,\quad 
\nonum
\Longleftrightarrow \quad 
\frac{1}{-\om V} \frac{1}{1 + \om/E_q} &\geq& \frac{1}{\lb}\,,
\quad q \neq 0\,.
\label{prop6}
\end{eqnarray} 
Since $\om <0$ both factors on the left hand side must be 
positive. Hence $E_q + \om >0$ for all $q\neq 0$, for which 
$\om > - 4 \sin^2 \pi/L$ is a necessary and sufficient condition. 

It is easy to sharpen this bound to
\be
\om >- \frac{4}{2d+1} \sin^2\frac{\pi}{L}
\label{ombound}
\end{equation}
by noting that 
\be
0>-\frac{V}{\lb}=\sum_p\frac{1} {E_p+\om}\ge 
\frac{1}{\om}+\frac{2d}{4\sin^2\frac{\pi}{L}+\om}\ .   
\end{equation}
From now on we write $\om_-(\lb,V) \in (-4 \sin^2 \pi/L, 0)$ for 
the unique root of the $\lb D(0) =-1$ gap equation such that 
$-\lb D_-(x) \geq 1$ for all $x$. Its large volume asymptotics comes 
out as 
\be 
- V \om_-(\lb, V) = \left\{ 
\begin{array}{ll} 
\displaystyle{\frac{4\pi}{\ln V}} 
\Big(1 + O(1/\ln V)\Big) \quad & d =2\,,
\\[4mm]
\Big(\dfrac{1}{\lb} + C_d \Big)^{-1} + O(V^{-\frac{d-2}{d}}) 
\quad & d\geq 3\,.   
\end{array}
\right.
\label{prop7}
\end{equation}
where $C_d = \int_0^{2\pi} \frac{d^d p}{(2\pi)^d}\,\frac{1}{E_p}$. 

In summary, to leading order the invariant spin two-point functions 
in the compact ferromagnetic and in the noncompact model can be written 
in the form (\ref{prop1}), where 
\be 
D_{\pm}(x) := D(x)|_{\om = \om_{\pm}(\lb,V)}\,,\quad \mbox{with}\quad 
\om_{\pm}(\lb,V) \;\;\mbox{solution of}\;\; \lb D(0) = \pm 1\,.
\label{prop8}
\end{equation}
In both the compact model and the noncompact model there is a 
unique  root of the gap equation $\lb D(0) = \pm 1$ 
such that $\lb D_{\pm}(x) \leq \pm 1$ for all $x$. 
From now on the symbols $D_{\pm}(x)$ will always refer 
to the unique propagators with this property. 
They satisfy
\ba 
&& 1 \leq - \lb D_-(x) < - 1 - \frac{2\lb}{V \om_-}\,,
\nonum
&& -1 + \frac{2 \lb}{V \om_+} < \lb D_+(x) \leq 1\,.
\label{prop9}
\end{eqnarray}
The $\om_{\pm}$ dependent bounds follows from $D(x) > 2/(V \om) - D(0)$. 
In particular $0< - \om_- < \lb/V$.   

For $V \ra \infty$, $D_+(x)$ approaches the massive free
propagator of squared mass $\om_+(\lb ,\infty)$, while $D_-(x)$
becomes massless and naively appears to be ill-defined. If one inserts, 
however, for $\om$ the $L$-dependent solution of the gap equation  $-\lb 
D_-(0) =1$ one can rewrite $\lb D_-(x)$ in such a way that the 
existence of the thermodynamic limit is manifest:
\be
\bra n_x \cdot n_y \ket\Big|_{N = \infty} = -\lb D_-(x-y)= 
1 + \frac{\lb}{V} \sum_{k \neq 0}\frac{1-\cos  k \!\cdot \!(x-y)} {E_k + 
\om_-}\, ,
\end{equation}
which for $V \ra \infty$ has the well-defined limit
\be
 1+\lb \int \! \frac{d^dk}{(2\pi)^d}\,\frac{1-\cos  k \!\cdot \!(x-y)}{E_k} \,.
\end{equation}


\newsubsection{Large $N$ correspondence} 

The results of sections 2 and 3 now lead to the 

{\bf Result:} Let $W_{r,+}$ and $W_{r,-}$ denote the connected parts of 
the invariant $2r$-point functions $\bra n_{x_1} \cdot n_{y_1} \ldots 
n_{x_r} \cdot n_{y_r} \ket$ in the ${\rm SO}(N+1)$ and in the 
${\rm SO}(1,N)$ nonlinear sigma-model, respectively. (With bare coupling 
$\beta$, defined on a hypercubic lattice $\Lambda \subset \Z^d$ of volume $L^d$ 
with periodic boundary conditions, and a suitable gauge fixing 
in the noncompact case.) Then the $W_{\pm,r}$ have well-defined 
``large $N$'' asymptotic expansions of the form (\ref{def5}),
with $\lb := (N+1)/\beta$ fixed. Further:  
\begin{itemize} 
\itemsep -3pt
\item[(a)] The coefficient functions $W_{\pm,r}$ are 
translation invariant. The lowest order two-point functions 
have the form $W_{\pm,1}^{(0)}(x,y) = \pm D_{\pm}(x,y) -1/\lb$, where 
$D_{\pm}(x) = D(x)|_{\om \ra \om_{\pm}}$. Here $D(x)$ is the free 
propagator of squared mass $\om$ and $\om_{\pm}(\lb,V)$
are the unique solutions of the gap equations $\pm \lb D(0) = 1$ 
with the property $\pm \lb D(x) \geq 1$, for all $x$. 
\item[(b)] For all $r \geq 1$, $s \geq 0$, there exist unique 
functionals $X_r^{(s)}[D](\lb)$ of a free propagator $D$ such that 
$W_{+,r}^{(s)} = X_r^{(s)}[D_+](\lb)$ are the coefficients in 
the compact model and $W_{-,r}^{(s)} = (-1)^r X_r^{(s)}[D_-](-\lb)$
are the coefficients in the noncompact model.   
\end{itemize} 

{\it Proof.} The existence of the large $N$ asymptotic expansion
has been shown in the compact model in \cite{kupi}, for $W[H]$
(or its the counterpart with a linear source coupling). 
Since $W_+^d[H] = W^d[H] = W[H]$, this directly carries 
over to $W_+[H]$. In the noncompact model the result
has been anticipated in Subsection 3.1; for the derivation we 
refer to the forthcoming paper \cite{horo}. 
The two statements in part (a) follow from sections 
2.1 and 3.3, respectively. For (b) we use the result stated 
in section 3.1, namely that $W_-^d[H]$ (although not equivalent 
to $W_-[H]$ and not obtainable from a well-defined duality 
transformation) generates the correct large $N$ asymptotic 
expansion. Given this, it suffices to note that 
$W_+^d[H]$ in (\ref{cd3}) and $W_-^d[H]$ in (\ref{wdnc}) 
have the same structure with actions that are related by the involution 
\be 
S_+[\alpha,H] \mapsto S_-[\alpha,H]\quad 
\mbox{for} \quad \alpha_x \mapsto - \alpha_x, \;\; \lb \mapsto -\lb\,.
\label{flip}
\end{equation}
Hence, if indeed the moments of $W_+^d[H]$ and 
$W_-^d[H]$ have asymptotic expansions of the form (\ref{def5}),
their coefficients are uniquely defined and must therefore also be related 
by the involution (\ref{flip}). 

For the sake of clarity let us add two remarks: 

First, $W_+[H]$ and $W_-[H]$ do {\it not} coincide exactly, as follows 
from (\ref{ineq}), (\ref{equ}) and the fact that the boundary $\cD(H)$ 
in (\ref{Wa3}) is invariant under the involution (\ref{flip}). In physics 
terms this amounts to the observation made before that the 
antiferromagnet and the noncompact model are inequivalent.   
 
Second, (b) does {\it not} state that the large $N$ coefficients 
of the compact (ferromagnetic or antiferromagnetic) model and the 
noncompact model are related by flipping the sign of $\lb$.  
The latter is certainly incorrect. Rather the dependence 
on $D_{\pm}$ is same after flipping the sign of $\lb$,
but $D_+$ and $D_-$ are very different and independently defined 
functions of the lattice distance. So, given the large $N$ coefficients 
of $W_{+,r}$, as a function of the lattice points, there would be 
no simple way to construct the large $N$ expansion of $W_{-,r}$.

\newsubsection{Schwinger-Dyson equations}

It is instructive to look at the previous result from the 
viewpoint of the Schwinger-Dyson (SD) equations. Under the assumption
that asymptotic expansions of $W_{-,r}$ and $W_{+,r}$ 
are known to exist, this also provides an alternative derivation 
of the result. 

In the compact model the Schwinger-Dyson equations for the
moments of $W[H]$ have, to our knowledge, first been formulated by 
M. L\"uscher \cite{luscher}.
In the noncompact model the necessity to gauge fix requires 
modifications. In the fixed spin gauge the modifications in the 
equations are minimal, but they inevitably refer to the preferred point 
$x_0$. In the following we formulate the SD equations in parallel for the 
gauge fixed functionals $W_{\pm}[H]$. 

The result is: 
\ba 
\label{sd1}
&& \pm \beta \Delta_z \Big[h_{zy} W_{\pm} - h_{zx} W_{\pm} - h_{zx} h_{xy} W_{\pm} 
- (h_{zx} W_{\pm})(h_{xy} W_{\pm}) \Big]\Big|_{z = x} 
\nonum
&& + \sum_{z \neq x} H_{xz} \Big[ h_{zy} W_{\pm} - h_{zx} W_{\pm} - h_{xy} W_{\pm} - 
h_{xz} h_{xy} W_{\pm} - (h_{zx} W_{\pm})(h_{xy} W_{\pm}) \Big]
\nonum
&& - N (1-\delta_{xy}) (h_{xy} W_{\pm} +1) =0\,.
\end{eqnarray} 
Again the upper sign refers to the compact model and the lower 
one to the noncompact model. In addition there is the following 
minor but essential modification: in the compact case 
(\ref{sd1}) holds for all $x,y \in \Lambda$, since $W_+[H]$ 
coincides identically with the generating functional $W[H]$ 
defined without gauge fixing, see (\ref{wc2}).
In contrast, in the noncompact case it  holds for all 
$x_0 \neq x \in \Lambda$ and all
$y \in \Lambda$. Specializing to $y = x_0$ thus results in an equation 
that is qualitatively different from the others. 

The derivation of (\ref{sd1}) is based on the invariance of the product 
measure $\prod_x d\Omega(n_x)$ under rotation of any one of the spins, 
$n_{x_1}$, say, where $x_1 \neq x_0$ in the noncompact model. 
For an infinitesimal rotation with the Lie algebra element $t^{ab}$ we 
write
\be 
\delta^{ab}_{x_1} n_x^d = \delta_{x_1 x}\, n_x^c (t^{ab})_c^{\;\;d}\,.
\label{sd2}
\end{equation} 
For later use we pick an explicit basis and note the completeness relations:
\begin{subeqnarray}  
\label{sd0}
so(N+1): \quad  
&& (t^{ab})_c^{\;\;d} = \delta_c^a \delta^{bd} - \delta_c^{\;\;b} \delta^{ad} 
= (t_{ab})_c^{\;\;d}\;,
\nonum
&& \frac{1}{2} \sum_{ab} (t^{ab})_c^{\;\;d} (t_{ab})_e^{\;\;f} =
\delta_{ce} \delta^{fd} - \delta_c^{\;\;f} \delta_e^{\;\;d}\,,
\\
so(1,N):\quad 
&& (t^{ab})_c^{\;\;d} = \delta_c^a \eta^{bd} - \delta_c^{\;\;b} \eta^{ad} 
= \eta^{aa'} \eta^{bb'} (t_{a'b'})_c^{\;\;d}\;,
\nonum
&& \frac{1}{2} \sum_{ab} (t^{ab})_c^{\;\;d} (t_{ab})_e^{\;\;f} =
\eta_{ce} \eta^{fd} - \delta_c^{\;\;f} \delta_e^{\;\;d}\,,
\end{subeqnarray}
where $\eta$ is defined by $a^c \eta_{cd} b^d = a^0 b^0 - 
a^1 b^1 - \ldots a^N b^N$. Denoting by $\bra \;\;\ket_H$ the source 
dependent averages this gives in a first step
\be 
\bra - (\delta_x^{ab} S) \cO + \sum_z H_{xz} t^{ab} n_x \cdot n_z\,
\cO + \delta_x^{ab} \cO\ket_H =0\,,
\label{sd3}  
\end{equation}
for any local function $\cO = \cO(\{n\})$ of the spins, and 
$x \neq x_0$ in the noncompact case. Specializing to $\cO = 
n_x \cdot t^{ab} n_y$ and summing over $a,b$, using the 
completeness relations (\ref{sd0}) gives 
\ba 
&& \Big\langle\!\! \pm \beta 
\Delta_z[n_z \cdot n_y - n_z \cdot n_x \; n_y \cdot n_x]\Big|_{z =x} 
+ \sum_{z \neq x} H_{xz} [n_z \cdot n_y - 
n_z \cdot n_x\; n_y \cdot n_x] 
\nonum
&& \sspace - N (1-\delta_{xy}) n_y \cdot n_x\Big\rangle_H =0 \,.
\label{sd4}
\end{eqnarray}    
Replacing the $\bra \;\; \ket_H$ averages by functional derivatives 
of $W_{\pm}[H]$ results in (\ref{sd1}).

Coupled bilinear equations for the $W_r$ functions arise by differentiating 
(\ref{sd1}) repeatedly and setting $H$ equal to zero. The equation 
obtained by $s$-fold differentiation involves all $W_r$ functions with
$r \leq s +2$; as a consequence no finite subsystem of equations 
arises for any individual $W_r$.

Before turning to the large N expansion of the system (\ref{sd1}) 
we wish to stress that even the exact SD equations do {\em not}
determine their solution uniquely. To see this, let us
momentarily denote the functional equations (\ref{sd1}) with the
upper and the lower sign by $(\rm{SD})_+$ and $(\rm{SD})_-$, 
respectively. Further we write $W_{+,\beta}$, $W_{+,-\beta}$, and
$W_{-,\beta}$, $\beta > 0$, for the generating functionals
of the ferromagnet, the anti-ferromagnet, and the noncompact
model, respectively. Then, by construction, $W_{+,\beta}$
solves $(\rm{SD})_+$ while both $W_{+, -\beta}$ and $W_{-,\beta}$
solve $(\rm{SD})_-$. (Consistency requires that substituting
the rhs of (\ref{flipW}) into the $(\rm{SD})_-$ equation converts it
back into a $(\rm{SD})_+$ equation for $W_{+,\beta}$ evaluated
on $H^{\eps}$. Using $(\Delta \eps f)_x = - \eps_x (\Delta f)_x 
- 4 d \eps_x f_x$, this can be verified to be the case.) Thus if
the exact SD equations were to determine their solution
uniquely, $W_{+,-\beta}$ and $W_{-,\beta}$ would have to
coincide. This however contradicts the discussion in section 2.2,
viz the antiferromagnet and the noncompact model are physically
and mathematically inequivalent. The upshot is that 
somehow initial or boundary conditions have to
be imposed on (\ref{sd1}) to specify a solution uniquely.
Since the exact equations couple all multipoint functions,
this seems difficult to do concretely.

In contrast, the large $N$ ansatz (\ref{def5}) effectively
converts the equations into ones which can be solved
recursively, and `initial' conditions can be specified.
The structure of the large $N$ expanded SD equations is
best explained by spelling out the first few (see \cite{TDlimit}) 
from which one can read off the recursion pattern for the 
$W_r^{(s)},\, r + s >1$, functions:

$$
\begin{array}{llll} W_1^{(0)} &\ra W_2^{(0)} &\ra W_3^{(0)} \ra \\[3mm]
\mbox{ } & \;\phantom{\ra} \downarrow & \phantom{\ra} \downarrow \\[3mm]
\mbox{} & \;\phantom{\ra} W_1^{(1)} & \ra W_2^{(1)} \ra \\[3mm] 
\mbox{} & \;\phantom{\ra} \mbox{}  & \phantom{\ra} \downarrow 
\end{array}
$$
\begin{center}
{\small Fig.~1: Recursion pattern for the solution of the large $N$ 
expanded SD equations.} 
\end{center}

To compute a given coefficient all quantities having 
arrows pointing towards it are needed. The detailed form 
of the equations and the solutions is not essential
for the following argument. 

We only need: (i) the fact that the equation for 
$W_1^{(0)}$ is autonomous and reads 
\ba
&& \pm \Delta_z[ W_1^{(0)}(z,y) - W_1^{(0)}(z,x)]\Big|_{z =x} 
\mp \lb W_1^{(0)}(x,y) \Delta_z W_1^{(0)}(z,x)\Big|_{z=x} 
\nonum
&& \sspace = 1 - \delta_{xy} + \lb W_1^{(0)}(x,y)\,,
\label{sd6}
\end{eqnarray} 
and (ii) the assumption that each recursion step in Fig.~1
has a unique solution. We first present the solutions of 
(\ref{sd6}) and then discuss the status of the assumption (ii). 

A solution of (\ref{sd6}) with the upper sign is 
\be 
W_1^{(0)}(x,y) = D_+(x-y) - 1/\lb  \bspace \mbox{ferromagnet} \,,
\label{sd6a}
\end{equation}
Two solutions of (\ref{sd6}) with the lower sign are 
\ba
W_1^{(0)}(x,y) \is \eps_x \eps_y D_+(x-y) - 1/\lb  
\sspace \quad \;\;\mbox{anti-ferromagnet} \,,
\nonum
W_1^{(0)}(x,y) \is - D_-(x-y) - 1/\lb  \bspace \mbox{noncompact model} \,,
\label{sd6b}
\ea
where for the verification of the first solution 
$(\Delta \eps f)_x = - \eps_x (\Delta f)_x - 4 d \eps_x f_x$ has been used.  
For each solution $W_1^{(0)}(x,x) =0$ amounts to the corresponding gap equation. 
As discussed in section 3.3 the gap equations have $O(V)$ solutions other 
than the physical ones $\om_{\pm}$ entering $D_{\pm}$. Each solution of the 
gap equation defines a different solution of (\ref{sd6}) via (\ref{sd6a}), 
(\ref{sd6b}), and in principle there could be others. The selection of a 
specific solution -- here $D_{\pm}$ -- can be justified either from the 
functional integral (in that it gives rise to valid asymptotic expansion) 
or from the physics one seeks to describe.

The uniqueness of the recursion (ii), to all orders, 
is probably difficult to establish directly from the SD equations. 
In a perturbative expansion of the $W_r$ the 
problem becomes linear and the fact that the perturbative 
expansions of the  $W_r$ is uniquely determined by the 
equations has been pointed out by M. L\"{u}scher \cite{luscher}.
In other words a potential nonuniqueness is known to be of order 
$O(\beta^{-\infty})$, directly from the SD equations.

An indirect way to establish the uniqueness (ii) is by 
showing that the moments of the generating functionals 
$W_{\pm}[H]$ have well-defined asymptotic expansions of the 
form (\ref{def5}). The expansion coefficients then must be 
unique, and since $W_{\pm}[H]$ are solutions of (\ref{sd1}) the 
coefficients $W_{\pm,r}^{(s)}$ must be 
solutions of the expanded SD equations, and the unique 
solutions. This also implies that the solutions must 
be translation invariant, despite the preferred role of 
the $x = x_0$ equation.  As mentioned earlier, the fact that 
the moments of $W_+[H]$ have an asymptotic expansion follows 
from \cite{kupi}; for $W_-[H]$ this will be shown in \cite{horo}

Given the uniqueness of the recursion one can readily 
re-derive the result of section 3.4. From (\ref{sd1}) 
and (\ref{def5}) it is clear that the involution 
\be 
\lb \mapsto - \lb\,,\quad W_r^{(n)}(\lb) \mapsto 
(-)^r \,W_r^{(n)}(-\lb)\,,
\label{sd7}
\end{equation} 
will map the Schwinger-Dyson equations of the compact model 
onto those of the noncompact model, to all orders of their 
large $N$ expansions. To draw conclusions about the solutions, 
however, the uniqueness asserted in (ii) is crucial. 
It implies that a sequence of solutions $W_r^{(n)}(\lb)$ in 
the compact model, based on a specific choice of $W_{1,+}^{(0)}(\lb)$, 
has a unique counterpart in the noncompact model based on the 
flipped initial solution $-W_{1,-}^{(0)}(-\lb)$ of the lower sign 
SD equations (\ref{sd6}). Since the physical solutions of the 
compact ferromagnet and the noncompact model are related in 
this way it follows that there exists unique functionals 
$X_r^{(s)}[D](\lb)$ such that $W_{+,r}^{(s)} = X_r^{(s)}[D_{+}](\lb)$ 
and $W_{-,r}^{(s)} = (-1)^r X_r^{(s)}[D_{-}](-\lb)$, where $D_{\pm}(x-y) 
:= \pm (W_{\pm,1}^{(0)}(x,y) +1/\lb)$ are the physical solutions 
of (\ref{sd6}). In summary, the previous discussion reproduces 
the parts (a),(b) of the result in section 3.4. 

We add some remarks. First, the involution (\ref{sd7}) is the 
counterpart of (\ref{flip}). Let us repeat that the involutions
(\ref{flip}) or (\ref{sd7}) do {\it not} imply a 
simple relation between the $W_r^{(n)}(\lb)$ and $W_r^{(n)}(-\lb)$ 
as functions of the lattice points, because already $W_1^{(0)}(\lb)$ and  
$W_1^{(0)}(-\lb)$ are based on solutions of different gap equations. 
It is rather the functional dependence on $D_\pm$ which is the same.

The involution also does {\it not} map the 
large $N$  coefficients of the compact model into those of 
the antiferromagnet, defined by $W_+[H]|_{\lb \ra -\lb}$.   
The reason is that the large $N$ saddle point relevant for the 
antiferromagnet is not the one obtained by reversing the sign of 
$\lambda$ in the gap equation. Rather is is given by the {\it same} 
equation as the saddle point for the ferromagnet, as is manifest from 
(\ref{sd6b}). Generally the large $N$ coefficients of the antiferromagnet 
are obtained simply by introducing appropriate products of the 
alternating sign function $\epsilon_x$ in those of the ferromagnet, 
as explained in Section 2.2 .

The special role of the SD equations (\ref{sd1}) for $y = x_0$ is 
related to the discussion in the paragraph following Eq. (\ref{transl}). 
It may suffice to illustrate the point with the lowest order equation 
(\ref{sd6}). The solution of the $y = x_0$ equation in (\ref{sd6}) is  
\be
\bra n_x^0 \ket_{f.s.} = \bar{n}_x + O\Big( \frac{1}{N+1}\Big) 
\quad \mbox{with} \quad \bar{n}_x := 1+ \lb W_1^{(0)}(x_0,x)\,.
\label{sd8}
\end{equation}      
The interpretation of this solution as the nonzero average of 
the $n_x^0$ component follows from \cite{DNS}.  
To higher orders the solutions of the $y=x_0$ equations 
produce corrections to (\ref{sd8}) or non invariant three-point 
functions, etc.

\newsubsection{Algorithm in fixed spin gauge}

Here we illustrate the main results from sections 3.1 (existence of the 
asymptotic expansion) and 3.4 (compact--non-compact correspondence) 
by an explicit computation of the two- and four-point functions in 
the noncompact model to next-to-leading order. The starting 
point is the generating functional (\ref{Wa2}), where according to 
the result described, $\cD(H)$ has been replaced by $\R^{V-1}$.   
Due to the gauge fixing a number of complications arise compared to the 
conventional large $N$ computations; in particular it is not obvious how  
translation invariant correlation functions are recovered.

Initially we keep the putative saddle point configuration 
$\om_x\,,x \neq x_0$, generic and specialize to $\om_x = \om_-$ 
only later. We thus write      
\be 
a_x = \om_x + \frac{u_x}{\sqrt{N+1}} \,,\quad u_x \in \R\,,\quad x \neq x_0\,.
\label{alg3}
\end{equation}
Although notationally cumbersome it is important to carefully 
distinguish ``hatted'', ``tilde'' and ``plain'' matrices. 
We set 
\ba 
&& M_{xy} := - \Delta_{xy} + \om_x \delta_{xy} =: 
\widetilde{M}_{xy} + \om_{x_0} \delta_{xx_0} \delta_{xy}\,, 
\quad \widetilde{D}_{xy} := (\widetilde{M}^{-1})_{xy}\,.
\nonum
&& U_{xy} := u_x \delta_{xy} =: \widetilde{U}_{xy} 
+ u_{x_0} \delta_{xx_0} \delta_{xy}\,. 
\label{alg4}
\end{eqnarray}
The matrices $M$ and $U$ are also defined for $x,y = x_0$. For $M$ we
assume $ \om_{x_0} \neq \om_- + \lb$, which ensures its invertibility; 
in $U_{xy}$ the variable $u_{x_0}$ is non-dynamical. 
Further we write $\widehat{M}$, $\widehat{U}$ for the matrices obtained from 
$M$, $U$, respectively, by deleting the $x_0$-th row and column and 
put $\widehat{D} = \widehat{M}^{-1}$. Then 
\begin{subeqnarray}
\label{alg5}
\widetilde{A} \is \widetilde{M} + \frac{1}{\sqrt{N+1}} 
\widetilde{U} + \frac{1}{N+1} \lb H\,, 
\nonum
\widetilde{A}^{-1} \is \Big( \1 + \frac{1}{\sqrt{N+1}} \widetilde{D} 
\widetilde{U} + \frac{1}{N+1} \lb \widetilde{D} H \Big)^{-1} \widetilde{D}\,,
\\
\widehat{A} \is \widehat{M} + \frac{1}{\sqrt{N+1}} \widehat{U} + 
\frac{1}{N+1} \lb \widehat{H}\,, 
\nonum
\ln \widehat{A} \is \ln \widehat{M} + 
\ln \Big( \1 + \frac{1}{\sqrt{N+1}} \widehat{D} \widehat{U} + 
\frac{1}{N+1} \lb \widehat{D} \widehat{H} \Big)\,.
\end{subeqnarray}
The required expansions are for square matrices $a,b$ 
\begin{subeqnarray}
\label{alg6} 
&& \Big( \1 + \frac{a}{\sqrt{N+1}} + \frac{b}{N+1} \Big)^{-1} 
= \1 + \sum_{l \geq 1} \frac{1}{(N+1)^{l/2}}\, q_l(a,b)\,,
\nonum
&& q_1(a,b) = -a\,,\quad q_2(a,b) = a^2 - b\,,\quad 
q_3(a,b) = - a^3 + a b + ba \,,
\nonum
&& q_4(a,b) = a^4 - a^2 b - a b a - b a^2 + b^2\,,
\\
&& \ln\Big( \1 + \frac{a}{\sqrt{N+1}} + \frac{b}{N+1} \Big) 
= \sum_{l \geq 1} \frac{1}{(N+1)^{l/2}}\, p_l(a,b)\,,
\nonum
&& p_1(a,b) = a\,,\quad p_2(a,b) = -\frac{1}{2} a^2 + b\,,\quad 
p_3(a,b) = \frac{1}{3} a^3 - \frac{1}{2}(a b + ba)\,,
\nonum
&& p_4(a,b) = - \frac{1}{4}a^4 + \frac{1}{3}( b a^2 + a b a + a^2 b) -
\frac{1}{2} b^2\,.
\end{subeqnarray}
Explicit formulas for the polynomials $q_l, p_l,\,l \geq 5$, can easily 
be found but will not be needed. With this notation one has 
\begin{subeqnarray}
\label{alg7}
(\widetilde{A}^{-1})_{x_0x_0} \is \widetilde{D}_{x_0x_0}
\Big[ 1 + \sum_{l \geq 1} 
\frac{1}{(N+1)^{l/2}}\, \widetilde{D}_{x_0x_0} Q_l \Big]\,,  
\\ 
Q_l &:=& \widetilde{D}_{x_0x_0}^{-2}\, 
\big(q_l(\widetilde{D}\widetilde{U}, 
\lb \widetilde{D}H)\widetilde{D}\big)_{x_0x_0} \,,
\nonumber
\\[1mm]
\ln \widehat{A} \is \ln \widehat{M} + 
\sum_{l \geq 1} 
\frac{1}{(N+1)^{l/2} }\,  p_l(\widehat{D} \widehat{U}, 
\lb \widehat{D} \widehat{H} ) \,,
\end{subeqnarray}
from which the expansion of $S[a,H]$, i.e.
\ba
S[a,H] \is \frac{1}{2} \Tr \ln \widehat{A} + \frac{1}{2\lb} \sum_{x \neq x_0} a_x 
- \frac{1}{2\lb} (\widetilde{A}^{-1})^{-1}_{x_0x_0}\,,
\nonum
S[a,H] \is \sum_{l \geq 0} \frac{1}{(N+1)^{l/2}} \,S_l[u,H]\,,
\label{aaction}
\end{eqnarray}
can readily be computed to any desired order. We present $S_1,\ldots, S_4$ first 
in the condensed form arising from (\ref{alg6}), (\ref{alg7}).  
\ba 
\label{alg8}
\nspace && S_1[u,H] = \frac{1}{2} \Tr p_1(\widehat{D} \widehat{U}, 
\lb \widehat{D} \widehat{H} ) + \frac{1}{2\lb} \sum_{x \neq x_0} u_x + 
\frac{1}{2\lb} Q_1\,,
\nonum
\nspace && S_2[u,H] =  \frac{1}{2} \Tr p_2(\widehat{D} \widehat{U}, 
\lb \widehat{D} \widehat{H} ) + \frac{1}{2\lb}\big(Q_2 -  
\widetilde{D}_{x_0x_0} Q_1^2\big)\,,  
\\
\nspace && S_3[u,H] = \frac{1}{2} \Tr p_3(\widehat{D} \widehat{U}, 
\lb \widehat{D} \widehat{H} ) + 
\frac{1}{2\lb}\big(Q_3   
- \widetilde{D}_{x_0x_0} 2 Q_2 Q_1 + \widetilde{D}_{x_0x_0}^2 Q_1^3\big)\,,  
\nonum
\nspace && S_4[u,H] = \frac{1}{2} \Tr p_4(\widehat{D} \widehat{U}, 
\lb \widehat{D} \widehat{H} ) +  
\frac{1}{2\lb}\big(Q_4 
- \widetilde{D}_{x_0x_0} (2 Q_3 Q_1 + Q_2^2) + 
3 \widetilde{D}_{x_0x_0}^2 Q_2 Q_1^2 - \widetilde{D}_{x_0x_0}^3 Q_1^4\big).  
\nonumber
\end{eqnarray} 
Note that $S_l$ is a homogeneous polynomial of order $l$ in the $u_x$'s and
$H$'s, assigning $u_x$ degree $1$ and $H_{xy}$ degree $2$.  

For $S_1$ and $S_2$ this gives in a first step 
\ba 
\label{alg9}
S_1[u,H] \is \frac{1}{2\lb}\sum_{x\neq x_0} u_x [ 1+ \lb \widehat{D}_{xx} - 
\widetilde{D}_{x_0x_0}^{-2} \widetilde{D}_{x_0 x}^2] \,,
\\[1mm]
S_2[u,H] \is \sum_{x,y \neq x_0} u_x u_y \bigg[ 
-\frac{1}{4} \widehat{D}_{xy}^2 + \frac{1}{2\lb} 
\frac{\widetilde{D}_{x_0 x}}{\widetilde{D}_{x_0x_0}} 
\widetilde{D}_{xy}
\frac{\widetilde{D}_{x_0 y}}{\widetilde{D}_{x_0x_0}} 
- \frac{1}{2\lb} \widetilde{D}_{x_0x_0} 
\frac{\widetilde{D}_{x_0 x}^2}{\widetilde{D}_{x_0x_0}^2}  
\frac{\widetilde{D}_{x_0 y}^2}{\widetilde{D}_{x_0x_0}^2} 
\bigg]
\nonum
&+&\frac{1}{2} \sum_{x,y} H_{xy} \bigg[ \lb \widehat{D}_{xy} - 
\frac{\widetilde{D}_{x_0 x}}{\widetilde{D}_{x_0x_0}} 
\frac{\widetilde{D}_{x_0 y}}{\widetilde{D}_{x_0x_0}} 
\bigg]\,.
\nonumber
\end{eqnarray} 
Using (\ref{d6}) one can now eliminate $\widetilde{D}_{xy}$ 
in favor of $D_{xy}:= (M^{-1})_{xy}$. This produces an expression of the 
same form just with $D_{xy}$ replacing $\widetilde{D}_{xy}$. 
Next we eliminate $\widehat{D}_{xy}$ in favor of $D_{xy}$, using 
(\ref{d2}). Anticipating $D_{x_0 x_0} = -1/\lb$ (which can be shown 
to hold independent of the gap equation) it follows that the 
saddle point equations $\delta S_1/\delta u_x =0$ are equivalent to   
\be 
- \lb D_{xx} = 1\,, \quad x \neq x_0\,,
\label{gap1}
\end{equation}
which is the non-translation invariant precursor of the gap equation used in 
section 3.3. Inserting this into $S_2$ gives 
\be 
S_2[u,H] =  - \frac{1}{4} \sum_{x,y\neq x_0} u_x u_y 
[D_{xy}^2 - \lb^2 D_{x x_0}^2 D_{y x_0}^2]
+ \frac{1}{2} \sum_{x,y} H_{xy} \lb D_{xy} \,.
\label{alg10}
\end{equation}

So far translation invariance was not presupposed, i.e.~we allowed 
for a nontrivial $x$-dependence in the extremal configurations 
$\om_x,\, x \neq x_0$.  We anticipate now  from 
\cite{horo} that the {\it only} solution of (\ref{gap1}) 
giving rise to a translation invariant  
$\bra n_x\! \cdot \!n_y \ket|_{N  = \infty} = - \lb D_{xy}$ 
is constant, i.e. 
\be 
\om_x = {\rm const},\, \quad x \neq x_0 \,.
\label{gap2}
\end{equation}
As seen in section 4.1 then $-\lb D(x) \geq 1$ fixes the constant to be 
$\om_- = \om_-(\lb,V)$, i.e. the unique solution of the 
translation invariant gap equation $-\lb D(0) = 1$ obeying 
$\om_- \in (-4 \sin^2 \pi/L, 0)$. As before we write $D_-(x-y)$ for 
the translation invariant propagator  with gap $\om_-$. 
Further, in the appendix we prove that 
\be 
S_2[u, H]\Big|_{\om_x = \om_-}  \geq 0  \,.
\end{equation}
This means that the saddle point is indeed a local 
minimum of the action (\ref{aaction}). 
We conjecture that it is in fact a global minimum of 
the action (\ref{aaction}) in the domain $\cD(H)$.

We now return to the generating functional $W[H]$. 
Substituting (\ref{alg10}) one arrives at 
\ba 
\label{alg11}
&& \exp W[H] = \exp\Big\{ - \frac{1}{2} \sum_{x,y} H_{xy} 
[\lb D_-(x-y) +1] \Big\} 
\nonum
&& \quad \times 
\cN\! \int \! \prod_{x \neq x_0} 
d u_x \exp\Big\{ + \frac{1}{4} \sum_{x,y\neq x_0} u_x u_y 
\widehat{D}_{-,2}(x,y) 
\Big\}
\nonum
&& \quad \times \bigg\{ 1 + \frac{1}{N+1}\Big(- S_4 + \frac{1}{2} S_3^2
\Big) + O\Big(\frac{1}{(N+1)^2}\Big) \bigg\}\,. 
\end{eqnarray} 
Here 
\be 
\widehat{D}_{-,2}(x,y) := D_-(x-y)^2 - \lb^2 D_-(x-x_0)^2 D_-(y-x_0)^2\,,
\end{equation} 
and the terms occurring in the third line are generated by the 
expansion of 
\be 
\label{alg12}
\exp\Big\{ - \sum_{l \geq 3} 
\frac{1}{(N+1)^{\frac{l-2}{2}}} S_l[u,H]\Big\}\,,
\end{equation}
with the half-integer powers of $1/(N\!+\!1)$ omitted. The conversion  
procedure 
exemplified for $S_2$ is easily seen to generalize to all $S_l,\, l\geq 3$. 
Indeed from $(\widetilde{A}^{-1})_{x_0x_0}^{-1} = 
-\om_{x_0} + (A^{-1})_{x_0x_0}^{-1}$ it follows that in an expansion 
\be 
\label{alg13}
(\widetilde{A}^{-1})_{x_0x_0}^{-1} = \widetilde{D}_{x_0x_0}^{-1} 
- \frac{1}{\sqrt{N+1}} Q_1 + \sum_{l \geq 2} \frac{1}{(N+1)^{l/2}}
\Big[ (\widetilde{A}^{-1})_{x_0x_0}^{-1} \Big]_l \,,
\end{equation}
all coefficients $[ (\widetilde{A}^{-1})_{x_0x_0}^{-1}]_l,\, l \geq 2$, 
have the following properties: (i) eliminating $\widetilde{D}_{xy}$ 
in favor of $D_{xy} = (M^{-1})_{xy}$ via (\ref{d6}), 
the parameter $\om$ drops out. (ii) The result of the replacement is 
an expression of identically the same form, just with $D_{xy}$ occurring 
for each instance of $\widetilde{D}_{xy}$. (iii) In the result one 
can replace $D_{xy}$ by $D_-(x-y)$ and $\widehat{D}_{xy}$ by 
$D_-(x-y) +\lb D_-(x-x_0) D_-(y - x_0)$.

The problem of computing $W[H]$ to order $1/(N\!+\!1)^p$
thus reduces to the algebraic problem of computing the $S_l[u,H]\,,l 
\leq 2 p + 2$, with the indicated substitutions, and to performing 
Wick contractions with the free propagator 
\be 
\label{alg14}
\bra u_x u_y \ket_0 = -2 \widehat{\Delta}_-(x,y)\,,
\sspace \sum_z \widehat{\Delta}_-(x,z) 
\widehat{D}_{-,2}(z,y) = \delta_{x,y}\,,\quad x,y \neq x_0\,.
\end{equation} 
In fact one has 
\be 
\widehat{\Delta}_-(x,y) = \Delta_-(x-y)\;,\sspace 
\mbox{for}\;\;x,y \neq x_0\,,
\label{alg15}
\end{equation}
where $\Delta_-$ is the translation invariant free $u$-propagator 
defined by $\sum_z \Delta_-(x-z) D_-(z-y)^2 = \delta_{xy}$.

On the basis of the general results the $H$-moments computed with the 
algorithm described above must be translation invariant and be related 
to that of the compact model by the involution described. 
As a test of both the algorithm and the involution we now compute the 
two- and the four-point function to sub-leading order. To this end 
explicit expressions for $S_3$ and $S_4$ are needed.

It is convenient to parameterize them as follows
\ba 
S_3 \is  S_3^{(0)} + S_3^{(1)} = \frac{1}{2} [\Tr p_3 + 
\lb Q_3^{(0)} + \lb^2 Q_3^{(1)}\,]\,,
\nonum
S_4 \is S_4^{(0)} + S_4^{(1)} + S_4^{(2)} = \frac{1}{2}[\Tr p_4 + 
\lb Q_4^{(0)} + \lb^2 Q_4^{(1)} + \lb^3 Q_4^{(2)}\,]\,.
\label{e18}
\end{eqnarray} 
Here $S_l^{(n)}$, $n=0,1,2$, are the pieces of order $n$ in the source $H$. 
The contribution to $S_l^{(n)}$ coming from 
$[ (\widetilde{A}^{-1})_{x_0x_0}^{-1}]_l$ in (\ref{alg13}) we denote 
by $Q_l^{(n)}$, for $n=0,1,2$, respectively. We need $Q_3^{(0)},\,Q_3^{(1)}$ 
and $Q_4^{(1)},\,Q_4^{(2)}$. They come out as 
\begin{subeqnarray}
\label{e19}
Q_3^{(0)} &=& -\!\! \sum_{x,y,z \neq x_0} u_x u_y u_z 
D(x\!-\!x_0) D(z\!-\!x_0) \dh(x,y) \dh(y,z)\,,  
\\
Q_3^{(1)} &=& 2 \sum_{x,y;z\neq x_0} H_{xy} u_z 
D(x\!-\!x_0) D(z \!-\!x_0) \dh(y,z)\,, 
\\
Q_4^{(1)} &=& -\!\! \sum_{x,y;z,w\neq x_0} H_{xy} u_z u_w 
D(z\!-\!x_0) \dh(w,x) 
\nonumber\\[-2mm] 
&& \quad \times [2 D(y\!-\!x_0) \dh(z,w) + D(w\!-\!x_0) \dh(z,y)]\,,
\\
Q_4^{(2)} &=& \sum_{x,y,z,w} H_{xy} H_{zw} D(x\!-\!x_0) D(w\!-\!x_0)
\dh(y,z)\,.  
\end{subeqnarray} 
Here we omitted the subscript `$-$' on the propagators and we shall 
continue to do so for the remainder of this subsection. 

For $S_3^{(0)}$, $S_3^{(1)}$, $S_4^{(1)}$, $S_4^{(2)}$ one obtains 
\begin{subeqnarray}
\label{e20}
S_3^{(0)} \is \frac{1}{6} \!\sum_{x,y,z \neq x_0} 
u_x u_y u_z \dh(x,y) \dh(y,z) 
[D(z\!-\!x) - 2 \lb D(x\!-\!x_0) D(z\!-\!x_0)]\,, 
\\
S_3^{(1)} \is - \frac{\lb}{2} \sum_{x,y;z \neq x_0} H_{xy} u_z 
\dh(x,z) [D(z\!-\!y) - \lb D(z\!-\!x_0) D(y\!-\!x_0) ]\,,
\\
S_4^{(1)} \is \frac{\lb}{4} 
\sum_{x,y;z,w \neq x_0} H_{xy} u_z u_w 
\dh(x,w) \Big\{ \dh(z,w) [D(z\!-\!y) -3 \lb D(z\!-\!x_0) D(y\!-\!x_0) ] 
\nonumber \\[-2mm]
&& \sspace + \dh(z,y) [D(w\!-\!z) - \lb D(w\!-\!x_0) D(z\!-\!x_0) ] \Big\} \,,
\\
S_4^{(2)} \is -\frac{\lb^2}{4}\sum_{x,y,z,w} H_{xy} H_{zw} 
\dh(y,z) [D(x\!-\!w) - \lb D(x\!-\!x_0) D(w\!-\!x_0)]\,. 
\end{subeqnarray} 
The desired correlation functions can now be computed 
from 
\ba 
\label{e21}
\bra -S_4^{(2)} + \frac{1}{2} [S_3^{(1)}]^2 \ket_0 
\is \frac{\lambda^2}{8} \sum_{x_1,y_1,x_2,y_2} H_{x_1,y_1} H_{x_2,y_2} 
W_2^{(0)}(x_1,y_1;x_2,y_2) \,,
\nonum
\bra -S_4^{(1)} + S_3^{(0)} S_3^{(1)} \ket_0 
\is \frac{\lambda}{2} \sum_{x,y} H_{xy} W_1^{(1)}(x,y)\,.
\end{eqnarray}
For $W_2^{(1)}$ the $x_0$-dependent terms cancel out algebraically 
and one finds 
\ba
\label{e22}
&& W_2^{(0)}(x_1,y_1;x_2,y_2) = 
D(x_1\! - \!x_2) D(y_1\! - \!y_2) + D(x_1 \!- \!y_2) D(y_1 \!- \!x_2) 
\nonum
&& \sspace - 2 \sum_{z,w} 
D(x_1\! - \!z) D(y_1 \!- \!z) \Delta(z\!-\!w) 
D(w \!- \!x_2) D(w \!- \!y_2)\,.
\end{eqnarray}
The expansion of $\bra -S_4^{(1)}\ket_0$ and  
$\bra S_3^{(0)} S_3^{(1)} \ket_0$ generates a large number 
of $x_0$-dependent terms which simplify after using 
$\sum_z \Delta(x-z) D(z-x_0)^2 = \delta_{xx_0}$. 
The results are 
\begin{subeqnarray} 
\label{e23}
&& \bra - S_4^{(1)} \ket_0 = \lb \sum_{xy} H_{xy} 
\sum_{z,w} \Delta(z,w) D(z\!-\!w) D(x\!-\!w) D(y\!-\!z) 
\\
&& \quad - \lb^3 \sum_{xy} H_{xy} D(x\!-\!x_0) D(y\!-\!x_0) 
\sum_{z,w} \Delta(z,w) D(z\!-\!w) D(z\!- \!x_0) D(w\!-\!x_0) \,.
\nonum
&& \bra S_3^{(0)} S_3^{(1)} \ket_0 = - \lb q 
\sum_{xy} H_{xy} \sum_z D(z\!-\!x) D(z\!-\!y) 
\\
&&  \quad + \lb^3 \sum_{xy} H_{xy} D(x\!-\!x_0) D(y\!-\!x_0) 
\sum_{z,w} \Delta(z,w) D(z\!-\!w) D(z\!- \!x_0) D(w\!-\!x_0) \,.
\nonumber
\end{subeqnarray} 
Here 
\be 
\label{e24}
q := \sum_z \Delta(z) \sum_{u,v} D(z\!-\!u) D(z\!-\!v) 
\Delta(u\!-\!v) D(u\!-\!v)\,.
\end{equation}
Finally 
\ba
\label{e25}
&& W_1^{(1)}(x,y) = -2 q \sum_w D(x\!-\!w) D(y\!-\!w) 
\nonum
&& \quad + 2 \sum_{z,w} D(x\!-\!z) D(y\!-\!w) 
\Delta(z\!-\!w) D(z\!-\!w)\,, 
\end{eqnarray} 
which obeys $W_1^{(1)}(x,x) =0$. 
Recall that $D(x)$ here is short for $D_-(x) = D(x)|_{\om = \om_-}$.
The results (\ref{e25}) and (\ref{e22}) are manifestly translation invariant 
and related to those of the compact model (see \cite{crist,TDlimit}) 
by the involution described.  

\newsection{Conclusions} 

On the level of asymptotic expansions interesting and useful correspondences
exist between compact  and noncompact nonlinear sigma-models. Here we 
established two such correspondences: in perturbation theory, where 
simply flipping the sign of $\beta$ allows to move between the compact and 
noncompact models, and in the large $N$ expansion, where one has to make a 
detour through the finite volume in order to establish a correspondence  
that involves not just flipping the sign of $\lb$ but also picking the 
physically appropriate solution of the gap equation.

Taking the sign flip prescription beyond perturbation theory leads from the 
compact model {\it not} to the noncompact one, but rather to another compact 
model, the antiferromagnet. 

In the literature \cite{bgs} a sign flip rule for
the large $N$ coefficients in a continuum formulation has
been proposed. By section 3.4 the rule is incorrect.
In \cite{fg} the much weaker claim has been made that
there exists a scheme in which the coefficients of the large
$N$ beta function in the compact and in the noncompact models
are related by the sign flip $\lb \ra -\lb$. This statement seems 
to be in agreement with our findings because it is based on
re-organizing the perturbation expansion into powers of $1/N$,
but it remains an open question whether the analytic continuation 
from positive to negative $\lb$ employed in \cite{fg} to go from the 
compact to the noncompact model is justified in the continuum.
\vspace{1cm}

{\it Acknowledgements:} We wish to thank A. Duncan and M. L\"{u}scher 
for discussions and correspondence. 

\newpage

\setcounter{section}{0} 
\newappendix{Positivity of $S_2$} 

Here we show that 
\be 
S_2[u, H]\Big|_{\om_x = \om_-}  \geq 0  \,,
\end{equation}
which guarantees that the saddle point on which the large $N$ expansion 
is based is indeed a local minimum of the action (\ref{aaction}). 

We begin by noting that the source-dependent term separately obeys
\be 
\frac{1}{2} \sum_{x,y} H_{xy} \lb D(x-y) \geq 0\,,
\end{equation} 
for all sources $H_{xy} \leq 0$, as $-\lb D(x-y) \geq 1$. It 
therefore suffices to consider $S_2[u,0]$. Next we introduce 
\be 
\Pi(p) := \sum_x e^{i p x} D(x)^2 = 
\frac{1}{V} \sum_k \frac{1}{(E_k + \om) (E_{k-p} + \om)} \,,
\label{p8}
\end{equation}
and claim that in terms of it a sufficient condition for $S_2[u, 0] \geq 0$ is 
\be 
\Pi(p)  < 0\,, \quad p \neq 0\,.
\label{p9}
\end{equation}
To see this we prepare 
\be 
\Pi(0) \geq \frac{V}{\lb^2}\,,
\label{p10}
\end{equation}
using once more $1 \leq -\lb D(x)$. Then we rewrite 
$S_2[u,0]$ in Fourier space 
\ba
\label{p11} 
S_2[u,0] \is  
\frac{1}{4 V^2} \sum_{p,q} u(p)^* \,\cS_{pq} \,u(q)\,,
\nonum
\cS_{00} \is - \Pi(0) [V - \lb^2 \Pi(0)]\,,
\\[2mm]
\cS_{0p} \is \lb^2 v_0 v_p \;, \sspace v_p := \Pi(p) e^{- i p x_0} \,, 
\nonum
\cS_{p q} \is - \delta_{pq} V \Pi(p) + \lb^2 v_p^* v_q\,,\quad p,q \neq 0\,. 
\nonumber
\end{eqnarray}
Subject to (\ref{p9}) and (\ref{p10}) the matrix $\cS$ has the form of 
a positive rank one perturbation of a positive diagonal matrix, which 
therefore must be itself positive. 

It remains to show (\ref{p9}). We rewrite $\Pi(p)$ in the form  
\ba 
\Pi(p) \is - \frac{1}{E_p + 2\om} \Big[ \frac{2}{\lb} + J(p) \Big]\,,
\sspace p \neq 0\,,
\nonum
J(p) \is \frac{1}{V} \sum_k 
\frac{E_k + E_{p-k} - E_p}{(E_k + \om)(E_{p-k} + \om)}\,,
\label{p12}
\end{eqnarray}
using the gap equation. Here $J(p)$ manifestly has a finite 
thermodynamic limit. Further we claim 
\be 
J(p) \geq 0 \quad \mbox{for all} \;p\,,
\label{p13}
\end{equation} 
which implies (\ref{p9}). 

The proof of (\ref{p13}) is based on a reorganization of the $d$-fold sum 
in (\ref{p12}) such that each term is positive on account of the 
Lemma below. We begin by noting that $J(p)$ is completely symmetric in 
all arguments $p = (p_1,\ldots, p_d)$, and that 
\ba
J((2\pi - p_1, \tilde{p}_1)) & =& J(p)\,,\quad \tilde{p}_\mu := (p_1, \ldots, 
p_{\mu-1}, p_{\mu +1}, \ldots ,p_d)\,.  
\label{p14}
\ea
Since $J(0) > 0$ trivially, it suffices to restrict attention
to momenta $p\neq 0$ with $0\le p_1\leq \ldots \leq p_d \leq \pi$.
Further, from 
\be
E_{p-k}+E_k-E_p=8\sum_\mu  \cos\frac{p_\mu}{2} \sin\frac{k_\mu}{2}
\sin\frac{(k-p)_\mu}{2}\,,
\label{p15}
\end{equation}
one sees that for $L$ even 
\be
J((\pi,\ldots , \pi))=0\,,\sspace  J((0, \pi,\ldots, \pi))>0\,.
\label{p16}
\end{equation}
Next we insert (\ref{p15}) into (\ref{p12}) to obtain 
\ba
\label{p17} 
J(p) \is \frac{1}{4} \sum_{\mu} \;
\frac{1}{L^{d-1}} \sum_{k_\nu,\,\nu \neq \mu} S(\tilde{k}_{\mu}|p)\,,
\nonum
S(\tilde{k}_{\mu}|p) \is \frac{1}{L} \sum_{k_\mu}
\frac{2 \cos \frac{p_{\mu}}{2} \sin \frac{k_\mu}{2} 
\sin \frac{k_{\mu} - p_{\mu}}{2}}%
{[\sin^2 \frac{k_\mu}{2} - \sin^2 \frac{\alpha}{2} + X_{\mu}]%
[\sin^2 \frac{k_\mu-p_\mu}{2} -\sin^2\frac{\alpha}{2} + Y_{\mu}]}\,,
\ea
where we set
\be
X_{\mu} := \sum_{\nu \neq \mu} \sin^2 \frac{k_\nu}{2} \,,
\quad 
Y_{\mu} := \sum_{\nu \neq \mu} \sin^2 \frac{k_\nu - p_\nu}{2} \,,
\quad 
2 \sin \frac{\alpha}{2}  := \sqrt{-\om}\,.
\end{equation}  
The result (\ref{p13}) now follows from the 

{\bf Lemma:} For all $L \geq 2$ and $q \in [0, \pi], \, q\in \frac{2\pi}{L} \Z$, 
\be 
S(q,\alpha,x,y) := \frac{1}{L} \sum_{n =0}^{L-1} 
\frac{2 \cos \frac{q}{2} \sin \frac{\pi n}{L} \sin (\frac{\pi n}{L} -\frac{q}{2})}%
{[\sin^2\frac{\pi n}{L} - \sin^2 \frac{\alpha}{2} + x]%
[\sin^2 (\frac{\pi n }{L} - \frac{q}{2}) -\sin^2\frac{\alpha}{2} + y]} \geq 0\,,
\end{equation}
for parameters $x, y \geq 0$ and $0< \alpha < \pi/L$.

In the application to (\ref{p17}) only the bound on $\alpha$ needs to 
be verified. From (\ref{ombound}) we have however 
\be 
\sin \frac{\alpha}{2} < 
\sqrt{\frac{4}{2d +1}} \sin\frac{\pi}{2 L}\,.
\end{equation}     

Proof of the Lemma. Trivially $S(0,\alpha,x,y) >0$ and $S(\pi,\alpha,x,y) =0$
for $L$ even. It thus suffices to consider $q \in (0,\pi)$.
We make use of the following summation formulas:
\be 
\label{sum1}
\frac{1}{L} \sum_{n=0}^{L-1} \frac{1}{\sin^2 \frac{\pi n}{2} + 
\sh^2 \frac{a}{2} } = \frac{2 {\rm coth} \frac{La}{2}}{\sh a}\,,
\quad a \neq 0\,,
\end{equation}
and for $q \in \frac{2\pi}{L} \Z$ and $a \neq b \neq 0$: 
\ba 
\label{sum2}
&& \frac{1}{L} \sum_{n=0}^{L-1} 
\frac{1}{[\sin^2 \frac{\pi n}{L} + \sh^2 \frac{a}{2}]
[\sin^2( \frac{\pi n}{L} - \frac{q}{2}) + \sh^2 \frac{b}{2}]} 
\\
&& \quad = \frac{2}{\sh \frac{La}{2} \,\sh \frac{Lb}{2} \,\sh a \,\sh b}
\Big[ 
\frac{\sh \frac{L}{2}(a+b)\, \sh(a+b)}{\ch(a+b) -\cos q} +
\frac{\sh \frac{L}{2}(a-b)\, \sh(a-b)}{\ch(a-b) -\cos q} \Big]\,.
\nonumber
\end{eqnarray}
For $a=b \neq 0$ the result of the summation is obtained from the 
rhs of (\ref{sum2}) by omitting the second term in square brackets and 
specializing the remainder to $a =b$ (i.e.~no pole occurs for $a=b$ and 
$q =0$). Both (\ref{sum1}) and (\ref{sum2}) can be obtained from 
\be 
\frac{1}{L} \sum_{n=0}^{L-1} \frac{1}{e^{\frac{2\pi i n}{L}} -a} = 
\frac{a^{L-1}}{1 - a^L}\,, \;\quad a^L \neq 1\,. 
\label{sum3}
\end{equation}

In the application to $S$ we set $\sh^2 \frac{a}{2} = 
x - \sin^2 \frac{\alpha}{2}$, 
$\sh^2 \frac{b}{2} = y - \sin^2 \frac{\alpha}{2}$, and pick the 
roots $a ,b >0$ for $x,y > \sin^2\frac{\alpha}{2}$. In a first step one 
obtains (also using the 1d version of (\ref{p15}) backwards) 
\ba 
\label{l1}
S &=& 
\frac{2{\rm coth}\frac{La}{2}}{\sh a}
+ \frac{2{\rm coth}\frac{Lb}{2}}{\sh b}
- \frac{2(\sh^2 \frac{a}{2} + \sh^2 \frac{b}{2} + \sin^2 \frac{q}{2})}%
{\sh \frac{La}{2} \,\sh \frac{Lb}{2} \,\sh a \,\sh b}
\nonum
&& \times 
\Big[ 
\frac{\sh \frac{L}{2}(a+b)\, \sh(a+b)}{\ch(a+b) -\cos q} +
\frac{\sh \frac{L}{2}(a-b)\, \sh(a-b)}{\ch(a-b) -\cos q} \Big]\,.
\ea
This can be rewritten in the form 
\be 
\label{l6}
S = \frac{\cos^2 \frac{q}{2}}{\ch \frac{a}{2} \ch \frac{b}{2} \,
\sh \frac{La}{2} \sh \frac{Lb}{2}} 
\Big[ \frac{\sh \frac{L}{2}(a+b) \,\sh \frac{a+b}{2} }%
{\sh^2 \frac{a+b}{2} + \sin^2 \frac{q}{2}}  
- \frac{\sh \frac{L}{2}(a-b) \,\sh \frac{a-b}{2} }%
{\sh^2 \frac{a-b}{2} + \sin^2 \frac{q}{2}} \Big]\,.  
\end{equation}
Since the function 
\be 
\label{l6b}
t(z) := \frac{\sh \frac{Lz}{2} \,\sh \frac{z}{2}}%
{\sh^2 \frac{z}{2} + \sin^2 \frac{q}{2}}\,,  
\end{equation}
is positive and monotonically increasing on $\R_+$ it follows that 
for $a,b>0$ and $q \in (0,\pi)$ the rhs of (\ref{l6}) is nonnegative.

Next consider the case where $0< \beta:= i b < \pi/L$ and $a >0$. 
The expression (\ref{l6}) can in this case be written as 
\ba 
S \is \frac{4 \cos^2 \frac{q}{2}}{\ch \frac{a}{2} \cos \frac{\beta}{2}}
\frac{{\rm coth} \frac{La}{2} \,s_1 - {\rm cot} \frac{L \beta}{2} \,s_2}%
{(\ch a \cos \beta - \cos q)^2 + \sh^2 a \sin^2 \beta}\,,
\nonum
s_1 \is \sh \frac{a}{2} \cos \frac{\beta}{2} [\ch a -\cos \beta + 1 -\cos q]\,,
\nonum
s_2 \is \ch \frac{a}{2} \sin \frac{\beta}{2} [\ch a -\cos \beta - 1 +\cos q]\,.
\label{l7}
\ea     
Further 
\ba 
\label{l8}
&& {\rm coth} \frac{La}{2} \,s_1 - {\rm cot} \frac{L \beta}{2} \,s_2
= (\ch a - \cos \beta)\, 
\ch \frac{a}{2} \cos \frac{\beta}{2} \Big[ {\rm coth} \frac{L a}{2} 
{\rm th} \frac{a}{2} - \cot \frac{L \beta}{2} \tan \frac{\beta}{2} \Big]
\nonum
&& \quad + (1 - \cos q) \Big[ {\rm coth}\frac{La}{2} \,\sh \frac{a}{2} 
\cos \frac{\beta}{2} + 
\cot \frac{L \beta}{2} \sin \frac{\beta}{2} \,\ch \frac{a}{2} \Big]\,. 
\ea
The second term in (\ref{l8}) is manifestly positive. For the 
first term we note that 
$\beta \mapsto \cot \frac{L \beta}{2} \tan \frac{\beta}{2}$
maps $[0, \pi/L]$ bijectively onto $[1/L, 0]$, while 
$a \mapsto  {\rm coth} \frac{L a}{2} {\rm th} \frac{a}{2}$
maps $[0, \infty)$ bijectively onto $[1/L, \infty)$. 
Hence 
\be 
{\rm coth} \frac{L a}{2} 
{\rm th} \frac{a}{2} - \cot \frac{L \beta}{2} \tan \frac{\beta}{2} 
\geq 0\,,
\end{equation}
which shows that $S$ in (\ref{l7}) is nonnegative. 

It remains to consider the case $0 < \beta := i b < \pi/L$, 
and $0 < \alpha := i a < \pi/L$, (which may differ from the 
$\alpha$ used in the formulation of the Lemma). Specialization 
of (\ref{l6}) gives 
\be 
\label{l9}
S = \frac{\cos^2 \frac{q}{2}}{\cos \frac{\alpha}{2} \cos \frac{\beta}{2} \,
\sin \frac{L\alpha}{2} \sin \frac{L\beta}{2}} 
\Big[ \frac{\sin \frac{L}{2}(\alpha+\beta) \,\sin \frac{\alpha+\beta}{2} }%
{\sin^2 \frac{q}{2} -\sin^2 \frac{\alpha+\beta}{2}}  
- \frac{\sin \frac{L}{2}(\alpha-\beta) \,\sin \frac{\alpha-\beta}{2} }%
{\sin^2 \frac{q}{2} -\sin^2 \frac{\alpha-\beta}{2} } \Big]\,.  
\end{equation}
Since $x \mapsto - t(ix)$ (with $t$ from (\ref{l6b})) is again positive 
and strictly increasing 
for $x \in [0, \pi/L]$ and $q \in (0, \pi)$, it follows that 
the rhs of (\ref{l9}) is non-negative for the range of $\alpha,\beta$ 
considered. This concludes the proof of the Lemma.


\end{document}